\documentclass[tighten]{aastex63}

\newcommand*\diff{\mathop{}\!\mathrm{d}}

\graphicspath{{./}{figures/}}

\shorttitle{Model and Parameter Estimation}
\shortauthors{The \textit{NICER} Team}

\begin{document}

\title{CONSTRAINING THE NEUTRON STAR MASS--RADIUS RELATION AND DENSE MATTER EQUATION OF STATE WITH \textit{NICER}. III. MODEL DESCRIPTION AND VERIFICATION OF PARAMETER ESTIMATION CODES}

\correspondingauthor{Slavko Bogdanov}
\email{slavko@astro.columbia.edu}

\author[0000-0002-9870-2742]{Slavko Bogdanov}
\affiliation{Columbia Astrophysics Laboratory, Columbia University, 550 West 120th Street, New York, NY, 10027, USA}

\author[0000-0001-6157-6722]{Alexander J.~Dittmann}
\affil{Department of Astronomy and Joint Space-Science Institute, University of Maryland, College Park, MD 20742-2421, USA}
\affil{Center for Computational Astrophysics, Flatiron Institute, 162 5th Avenue, New York, NY 10010, USA}

\author[0000-0002-6089-6836]{Wynn C.~G.~Ho}
\affil{Department of Physics and Astronomy, Haverford College, 370 Lancaster Avenue, Haverford, PA 19041, USA}

\author[0000-0002-3862-7402]{Frederick K.~Lamb}
\affil{Illinois Center for Advanced Studies of the Universe and Department of Physics, University of Illinois at Urbana-Champaign, 1110 West Green Street, Urbana, IL 61801-3080, USA}
\affil{Department of Astronomy, University of Illinois at Urbana-Champaign, 1002 West Green Street, Urbana, IL 61801-3074, USA}

\author{Simin Mahmoodifar}
\affiliation{Astrophysics Science Division and Joint Space-Science Institute, NASA Goddard Space Flight Center, Greenbelt, MD 20771, USA}

\author[0000-0002-2666-728X]{M.~Coleman Miller}
\affiliation{Department of Astronomy and Joint Space-Science Institute, University of Maryland, College Park, MD 20742-2421, USA}

\author[0000-0003-4357-0575]{Sharon M.~Morsink}
\affiliation{Department of Physics, University of Alberta, Edmonton, AB T6G 2G7, Canada}

\author[0000-0001-9313-0493]{Thomas E.~Riley}
\affiliation{Anton Pannekoek Institute for Astronomy, University of Amsterdam, Science Park 904, 1090GE Amsterdam, The Netherlands}

\author[0000-0001-7681-5845]{Tod E.~Strohmayer}
\affiliation{Astrophysics Science Division and Joint Space-Science Institute, NASA Goddard Space Flight Center, Greenbelt, MD 20771, USA}

\author[0000-0002-1009-2354]{Anna L.~Watts}
\affiliation{Anton Pannekoek Institute for Astronomy, University of Amsterdam, Science Park 904, 1090GE Amsterdam, The Netherlands}

\author[0000-0002-2651-5286]{Devarshi Choudhury}
\affiliation{Anton Pannekoek Institute for Astronomy, University of Amsterdam, Science Park 904, 1090GE Amsterdam, The Netherlands}

\author[0000-0002-6449-106X]{Sebastien Guillot}
\affil{IRAP, CNRS, 9 avenue du Colonel Roche, BP 44346, F-31028 Toulouse Cedex 4, France}
\affil{Universit\'{e} de Toulouse, CNES, UPS-OMP, F-31028 Toulouse, France}

\author[0000-0001-6119-859X]{Alice K.~Harding}
\affil{Theoretical Division, Los Alamos National Laboratory, Los Alamos, NM 58545}

\author[0000-0002-5297-5278]{Paul S.~Ray} 
\affiliation{Space Science Division, U.S.~Naval Research Laboratory, Washington, DC 20375, USA}

\author[0000-0002-9249-0515]{Zorawar Wadiasingh}
\affil{Astrophysics Science Division, NASA Goddard Space Flight Center, Greenbelt, MD 20771, USA}
\affil{Universities Space Research Association (USRA), Columbia, Maryland 21046, USA}

\author[0000-0002-4013-5650]{Michael T.~Wolff}
\affiliation{Space Science Division, U.S.~Naval Research Laboratory, Washington, DC 20375, USA}

\author[0000-0001-9803-3879]{Craig B.~Markwardt}
\affiliation{X-Ray Astrophysics Laboratory, NASA Goddard Space Flight Center, Greenbelt, MD 20771, USA}

\author{Zaven Arzoumanian}
\affiliation{X-Ray Astrophysics Laboratory, NASA Goddard Space Flight Center, Greenbelt, MD 20771, USA}

\author{Keith C.~Gendreau} 
\affiliation{X-Ray Astrophysics Laboratory, NASA Goddard Space Flight Center, Greenbelt, MD 20771, USA}

\begin{abstract}
We describe the X-ray pulse profile models we use, and how we use them, to analyze \textit{Neutron Star Interior Composition Explorer} (\textit{NICER}) observations of rotation-powered millisecond pulsars to obtain information about the mass-radius relation of neutron stars and the equation of state of the dense matter in their cores. Here we detail our modeling of the observed profile of PSR J0030+0451 that we analyzed in \cite{miller19} and \cite{riley19} and describe a cross-verification of computations of the pulse profiles of a star with $R/M \sim 3$, in case stars this compact need to be considered in future analyses. We also present our early cross-verification efforts of the parameter estimation procedures used by \cite{miller19} and \cite{riley19} by analyzing two distinct synthetic data sets. Both codes yielded credible regions in the mass-radius plane that are statistically consistent with one another and both gave posterior distributions for model parameter values consistent with the values that were used to generate the data. We also summarize the additional tests of the parameter estimation procedure of \cite{miller19} that used synthetic pulse profiles and the \textit{NICER} pulse profile of PSR J0030$+$0451. We then illustrate how the precision of mass and radius estimates depends on the pulsar's spin rate and the size of its hot spot by analyzing four different synthetic pulse profiles. Finally, we assess possible sources of systematic error in these estimates made using this technique, some of which may warrant further investigation.
\end{abstract}


\section{Introduction} \label{sec:intro}

Strongly degenerate matter at densities above the density of nuclear saturation ($\rho_s \equiv 2.4\times10^{14}$\,g\,cm$^{-3}$) can exist stably in the cores of neutron stars (NSs). This makes NSs important natural laboratories for studying the physics of the strong interaction and the state of matter at supranuclear densities, especially because at present experimental and theoretical approaches to understanding dense matter suffer from multiple challenges (see, e.g., \citealt{2016RvMP...88b1001W} for a review).
A number of theoretically well-motivated proposals for the composition and properties of matter in the low-temperature, high-density regime are consistent with current data; they range from mainly nucleonic matter to matter with hyperons, deconfined quarks, color superconducting phases, or Bose-Einstein condensates \citep[see, e.g.,][]{lattimer16,Oertel17,Baym18}.  
Because the mass-radius relation of neutron stars depends sensitively on the pressure-density relation of the matter in their interiors \citep[see, e.g.,][]{Lattimer01,Lattimer05,Ozel09,Read09a,hebeler13,2013ApJ...765L...5S}, measurements of the mass and radius of several neutron stars to a precision of a few percent can, in principle, provide insights into the composition and properties of the matter in their cores, especially if the stars have different masses. This has prompted the development of various astrophysical techniques for inferring the masses and radii of neutron stars. Several such methods use observations of thermal electromagnetic radiation from the surface of the neutron star, which falls in the X-ray band when the temperature of the surface is $\gtrsim 0.1~{\rm keV}$ (see, e.g., \citealt{ozel13,miller13,heinke13,potekhin14,miller16}; and references therein). Detailed analyses of the pulse profiles of the pulsed X-ray emission produced by ``hot spots'' on the surfaces of rotating neutron stars have been identified as a particularly promising approach for determining the mass-radius relation of NSs. This technique has been studied extensively over the past four decades \citep{pechenick83,strohmayer92,1995ApJ...442..273P,pavlov97,miller98,braje00,weinberg01,beloborodov02,poutanen03,cadeau07,morsink07,lo13,psaltis14b,2014ApJ...791...78A,miller15,2018A&A...615A..50N}.

One of the primary scientific goals of the \textit{Neutron Star Interior Composition Explorer} (\textit{NICER}) mission \citep{2016SPIE.9905E..1HG} is to make precise and reliable measurements of the radii and masses of several neutron stars. To accomplish this, \textit{NICER} is conducting extensive observations of a promising set of nearby rotation-powered millisecond pulsars (MSPs) that exhibit thermal X-ray pulsations.  This paper is the third in a series of methodological papers describing how this approach can be used to measure the neutron star mass-radius relation and the dense matter equation of state using \textit{NICER}. \citet[][Paper~I hereafter]{bogdanov19a} described the \textit{NICER} observations and the resulting MSP data that are being collected, while \citet[][Paper~II hereafter]{bogdanov19b} described the approach and codes used to compute the pulse profile models that are being compared with the data being collected by \textit{NICER}. The purpose of this present work (Paper~III) is (i) to provide a compilation of the details of the pulse profile modeling technique that was applied to the \textit{NICER} data on PSR~J0030$+$0451 by~\cite{miller19} and~\cite{riley19} and that will be used in future analyses of pulse profile data obtained with \textit{NICER}; (ii) to describe the tests and cross-comparisons we have performed to check the parameter-estimation procedures that were used in~\cite{miller19} and~\cite{riley19} and will be employed in future work; (iii) provide an illustrative comparison between the mass-radius constraints that can be obtained for different assumed spins and hot spot sizes. and (iv) to provide a brief assessment of the possible systematic errors in this approach. The mass-radius posterior distributions for PSR~J0030$+$0451 that were presented in \citet{miller19} and \cite{riley19} were computed using these codes and the pulse profile models described here. \citet{miller19} and \citet{raaijmakers19,raaijmakers20} then used these posteriors to obtain constraints on the equation of state (EoS) of cold, dense matter.

The paper is organized as follows. In Section~\ref{sec:model}, we detail our modeling of the X-ray emission from the surface of a rapidly rotating neutron star and the resulting energy-resolved X-ray pulse profile models we use to constrain the mass-radius relation of neutron stars and the equation of state of the dense matter in their interiors. This section also reports a comparison of two independent computations of the pulse profile produced by a star with $R_{\rm eq}/M \approx 3$, performed to cross-check the accuracy of these codes in case stars this compact need to be considered in future analyses of \textit{NICER} pulse profile data. In Section~\ref{sec:verification}, we describe the tests and cross-comparisons we have performed to verify the two different parameter-estimation procedures presented in \cite{miller19} and \cite{riley19}. The first two tests we describe in detail here were carried out early in our development of parameter estimation procedures. This section also summarizes the results of two additional tests of the parameter estimation procedure used by \cite{miller19}, using a synthetic pulse profile that was designed to mimic the pulse profile of PSR~J0030$+$0451 observed with \textit{NICER}.  In Section~\ref{sec:model_properties}, we demonstrate how the neutron star's spin rate and the size of its hot spot affect the constraints on the star's mass and radius that can be obtained using the pulse-profile fitting method using synthetic data. In Section~\ref{sec:systematics}, we briefly discuss and assess several possible sources of systematic error in estimates of stellar masses and radii made using this technique and evaluate whether these warrant further investigation. We summarize the contents and discussion in this paper in Section~\ref{sec:discussion}.

\section{Model Components}\label{sec:model}

There are numerous physical processes that determine the characteristics of an  X-ray pulse profile observed far from an X-ray--emitting, rotating neutron star (NS). These include the generation of the X-rays that emerge from the star's surface, the propagation of this radiation in the exterior spacetime of the NS and then through the interstellar medium between the star and Earth, and the interaction of the X-rays with the optics and detectors of the observing telescope. In more detail, the effects that must be considered are:

\begin{enumerate}
\item  The physical properties of the emitting atmosphere of the star (including its density, chemical composition, ionization state, and magnetization) and the spectrum and beaming pattern of the radiation that emerges from each point on the star;
\item  The temperature distribution on the stellar surface, which we model as spots with various locations, sizes, shapes, and temperatures, each of which emits radiation that contributes to the total flux potentially observed using \textit{NICER} as a function of the star's rotational phase;
\item  The special relativistic boost and aberration of the emitted radiation caused by the rotational velocity of the emitting gas;
\item  The general relativistic gravitational lensing of the radiation emitted by the NS surface;
\item  The partial or complete occultation of each hot spot by the body of the star as it rotates;
\item  The general relativistic redshift and time dilation that affects the frequency and travel time of the radiation as it propagates from the emitting surface to the observer;
\item The energy-dependent attenuation of the emitted X-rays by the interstellar medium between the star and the observer;
\item The propagation of the incident X-ray photons through the \textit{NICER} concentrator optics and the efficiency of the detectors in registering them, both as functions of energy, information that is encapsulated in the redistribution matrix file (RMF) and the ancillary response file (ARF);
\item  The combined backgrounds produced by emission from the non-spot portions of the star, from other parts of the pulsar system (including any wind nebula, and/or binary companion), from neighboring astronomical sources, and from any diffuse emission in the \textit{NICER} field of view, as well as the background produced by the radiation environment of the telescope.
\end{enumerate}

We now describe each of these components of the model in more detail, including how they are incorporated in our analyses of \textit{NICER} data on MSPs. Items 3--6 were discussed in depth in Paper~II.

\subsection{Neutron Star Model Atmospheres} \label{sec:atmosphere}

\begin{figure}
\centering
\includegraphics[width=0.45\textwidth]{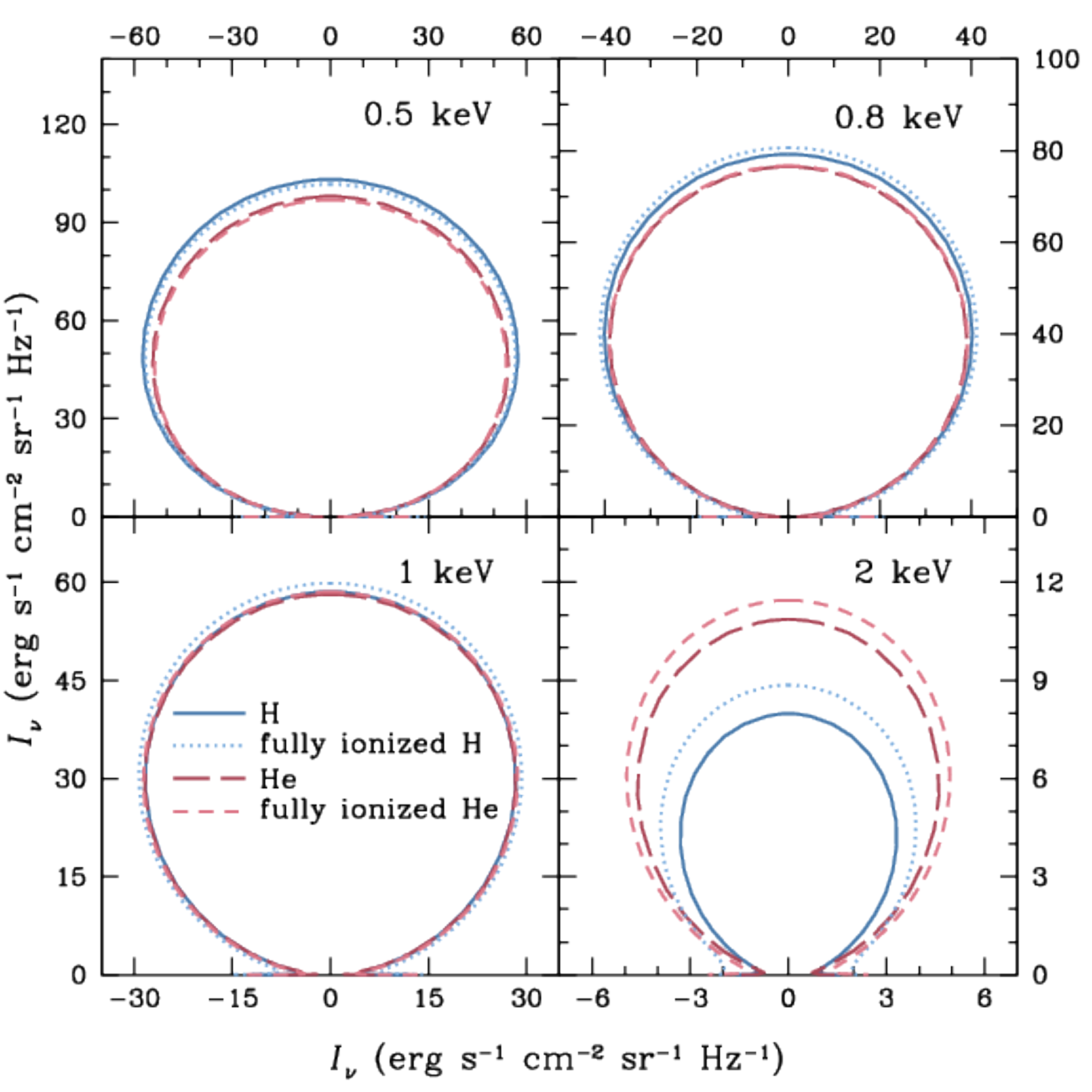}
\caption{
Polar diagrams of the specific intensity $I_\nu$ produced by
partially-ionized hydrogen (solid), fully-ionized hydrogen (dotted),
partially-ionized helium (long-dashed), and fully-ionized helium (short-dashed) unmagnetized gatmosphere models at the four photon energies indicated (0.5, 0.8, 1.0, and 2.0~keV), computed using the NSX code assuming
$T_{\rm eff}=10^6$~K and $g_s=2.4 \times 10^{14}$~cm~s$^{-2}$.
}
\label{fig:beam}
\end{figure}

The thermal X-rays observed from rotation-powered MSPs most likely originate in heated regions of a geometrically-thin but optically thick gaseous atmosphere that covers the stellar surface. The properties of this atmosphere, such as the strength of the magnetic field within it, its chemical composition, and its ionization state, directly determine the spectrum and beaming pattern of the radiation that emerges from the atmosphere (see, e.g., \citealt{potekhin14}, for a review). 

\textit{Magnetic fields}.---The characteristic strength at which the magnetic field in a neutron star atmosphere has a large effect on the radiation that emerges from it is $B_0 \equiv e^3m_{\rm e}^2 c/\hbar^3=2\times 10^9\mbox{ G}$ (see \citealt{lai01} for a review). At field strengths at or in excess of $B_0$, the electrons in the atoms and molecules in the atmosphere are more tightly bound to their nuclei than in weaker fields, their emission is strongly anisotropic and polarized (see, e.g., \citealt{meszaros92}), and spectral features produced by the electron cyclotron resonance and electron bound states may be visible in the spectrum of the radiation that emerges from the atmosphere (see, e.g., \citealt{lai01,2010A&A...518A..24P, 2010ApJ...714..630S, 2012ApJ...751...15S,potekhin14}). 

The external magnetic fields of rotation-powered pulsars have most commonly been assumed to be dipolar, with the dipole moment at the center of the star. With this assumption, the pulsar's magnetic dipole moment can be estimated from its pulse period $P$ and period derivative $\dot{P}$, using the braking formula for a force-free dipolar magnetic field\footnote{Studies of the radio and $\gamma$-ray emission of $\gamma$-ray pulsars have provided increasing evidence that their magnetic field structure resembles that predicted by force-free models; see \citet{2019ApJ...883L...4K} and references therein.} (see, e.g., \citealt{1999ApJ...511..351C}; \citealt{2006ApJ...643.1139C}; \citealt{2006ApJ...648L..51S}). For some of the MSPs of interest to us, the strength of the surface magnetic field inferred by making these assumptions is $\sim 5 \times 10^8~{\rm G}$, and hence significantly smaller than $B_0$ (see \citealt{miller19}). Thus, if the magnetic fields of these MSPs were centered dipoles, the appropriate force-free braking formula indicates surface fields that
would have only a small effect on the radiation produced there, and spectra and beaming patterns computed using non-magnetic model atmospheres (such as those shown in Figure~\ref{fig:beam}) would be adequate.

The preferred configuration of the temperature distribution for PSR\,J0030$+$0451 found by the independent analyses of \citet{miller19} and \citet{riley19} consists of one highly elongated (in the longitudinal direction) region and a second smaller, less elongated region, both on the hemisphere below the stellar equator relative to the observer's line of sight.  Such a temperature distribution is difficult to reconcile with a standard dipole field, but could be reproduced by assuming a dipole+quadrupole field\footnote{The implications of such a field geometry for different aspects of pulsars are discussed in \citet{bilous19} and \citet{miller19}.} (see, e.g., \citealt{Gralla2017,Lockhart2019,chen2020,2021ApJ...907...63K}). In such a case, the strength of the surface magnetic field inferred using a magnetic dipole braking formula could substantially underestimate the strength of the magnetic field within the atmospheres of hot spots (see \citealt{miller19}). Even if the star's external magnetic field is predominantly dipolar, if it is off-center, the dipole component of the magnetic field at the stellar surface could be $\sim 5$--10 times larger than would be inferred from the centered-dipole braking formula if the same magnetic field were centered in the star.  In that case the strength of the surface magnetic field near the magnetic poles could be larger than $B_0$. The absence of any obvious narrow-band features such as absorption lines in the spectrum of PSR~J0030$+$0415 observed using \textit{XMM-Newton} (see, e.g., \citealt{bogdanov09}) and in the $\sim 0.3$--3~keV spectrum observed using \textit{NICER} argues against surface magnetic fields with strengths $B \approx (\textrm{3--30}) \times 10^{10}$~G (assuming a gravitational redshift at the stellar surface of 20\%). While electron cyclotron features produced by much stronger magnetic fields $B \gg 10^{11}$~G would appear at $E > 3$~keV and therefore could be missed by these observations, such strong magnetic fields would create pencil or fan-beam emission, which would probably produce markedly different pulse profiles than the X-ray pulse profile observed using \textit{NICER}. Comparisons between beaming patterns produced by non-magnetic and magnetic hydrogen atmospheres with an effective temperature of $10^6$\,K show differences much greater than the differences between hydrogen and helium (see Figure~\ref{fig:beam}).

As discussed by \citet{miller19} and \cite{riley19}, energy-resolved pulse profile models constructed by assuming negligible magnetic fields in the atmospheres of the hot spots are consistent with the energy-resolved pulse profiles observed using \textit{NICER}. At present, there is no evidence that the available data require that the strengths of the magnetic fields in the atmospheres of the hot spots approach or exceed $B_0$, but the possibility of subtle effects introduced by strong fields may need to be examined (see Section~\ref{sec:systematics}).

\textit{Chemical composition}.---During their accretion-powered, spin-up phases, the chemical composition of the atmospheres of MSPs is expected to be determined primarily by the composition of the gas transferred from the envelope of the companion star. In many cases, the transferred gas is expected to be primarily hydrogen. Even if helium or heavier elements dominate in the accreted gas, some hydrogen is expected to be created during the accretion phase by spallation \citep{{1992ApJ...384..143B},{2019ApJ...887..100R}}.

The rotation-powered MSPs considered here are not expected to be actively accreting. Once accretion has ended, the heavier elements in their outer layers are expected to sink rapidly via diffusive gravitational separation, leaving behind the lightest element \citep{alcockillarionov80, hameuryetal83, Brown02}. The upper atmospheres of rotation-powered MSPs are therefore expected to consist of the lightest element that has been accreted, which in most cases is expected to be hydrogen. Accretion of even a small amount of hydrogen after mass transfer from the companoin has ended would also be sufficient to create an upper atmosphere of pure hydrogen \citep{blaesetal92, wijngaardenetal19, wijngaardenetal20}.

The upper atmosphere of a rotation-powered MSP could be helium if the hydrogen originally present in its atmosphere has all been converted to helium via diffusive nuclear burning \citep{Rosen68, changbildsten03, 2004ApJ...605..830C, wijngaardenetal19}, or if it accreted pure helium from a helium white dwarf companion during its mass-transfer phase. An MSP could have an atmosphere with even heavier elements, if no hydrogen or helium was ever accreted, if all the accreted hydrogen and helium has been burned to heavier elements by nuclear reactions, or if it has had a wind that has excavated its surface to expose underlying heavy elements \citep{2004ApJ...605..830C}. A comparison between beaming patterns produced by a helium atmosphere and a hydrogen atmosphere with an effective temperature of $10^6$\,K shows that they differ by $<5\%$ at $0.5-1$\,keV and $<30\%$ at 2\,keV (see Figure~\ref{fig:beam}). 

\textit{Ionization state}.---The opacity tables that are currently available for partially ionized hydrogen and helium \citep{iglesiasrogers96, badnelletal05, colganetal16} do not cover the entire range of energies and temperatures needed to model the pulse profiles of the MSPs we consider. However, the neutral fraction is low in atmospheres with the inferred temperatures ($\approx 0.5$--2$\times 10^6$~K) of the hot spots in these MSPs, and a comparison between the beaming patterns produced by a partially ionized atmosphere constructed using OP\footnote{The Opacity Project. See \url{http://cdsweb.u-strasbg.fr/topbase/OpacityTables.html}.} opacities \citep{badnelletal05} and a fully ionized atmosphere with an effective temperature of $10^6$~K shows that they differ by $<2$\% at 0.5--1~keV and $<11$\% at 2~keV (see Figure~\ref{fig:beam}). We have therefore used fully ionized model atmospheres in our work to date.

To model the pulse profiles and phase-resolved spectra of MSPs, one needs to determine the specific intensity $I_\nu(\mu)$ from a local patch of the NS surface, where $\nu$ is frequency of the radiation and $\mu=\cos\theta$ in terms of the angle $\theta$ between the direction of propagation of the radiation and the local normal to the surface. $I_\nu(\mu)$ is computed by solving the radiative transfer equation using established techniques \citep{mihalas78}. Early efforts to compute realistic spectra considered emission from (non-magnetic) fully-ionized, light-element atmospheres and anisotropic beaming patterns that depended on the chemical composition of the atmosphere and the photon energy \citep{romani87, rajagopalromani96, zavlinetal96}.  Since then, many non-magnetic NS model atmospheres have been constructed \citep{gansickeetal02,mcclintocketal04,heinkeetal06,hoheinke09,haakonsenetal12,suleimanovetal14,gonzalezcaniulef19}. For the parameter estimation analyses of the \textit{NICER} data presented in \citet{miller19} and \citet{riley19}, we used the NSX fully ionized H-atmosphere models of \citet{holai01} and \citet{hoheinke09}. For comparison, we also computed fully ionized He-atmosphere models. We plan to use these models in future analyses, to explore the sensitivity of the inferred $M$ and $R$ credible regions to the assumed chemical composition, among other things. In Section~\ref{sec:systematics}, we discuss the possible systematic errors that may be introduced by our assumption of a hydrogen atmosphere with a negligible magnetic field.

We note that non-local thermodynamic equilibrium (NLTE) effects \citep[see, e.g.,][]{rauch08} within the atmosphere are not important at temperatures and energies relevant for rotation-powered MSPs ($\sim 10^6$\,K and $\lesssim 1$\,keV). The same holds true for Compton scattering, which only becomes important at temperatures $\gtrsim 3\times 10^6$ K; deviations at lower temperatures only start to become noticeable at photon energies $E>2$\,keV but are still much less than 1\% and therefore would have no significant impact on our analysis \citep{Suleimanov2007,Salmi19}, given that nearly all the thermal emission observed from the MSPs observed by \textit{NICER} occurs below $\sim$2\,keV.

\textit{Computational method}.---Determining the specific intensity $I(E,\theta)$ emerging from an atmosphere as a function of the photon energy $E$ and emission angle $\theta$ is computationally expensive. For this reason, in our analyses of the \textit{NICER} pulse profile data we use pre-computed, high-resolution look-up tables of $I(E, \theta)$ for grids in the logarithm of the effective temperature, $\log T_{\rm eff}$, the logarithm of the surface gravity, $\log g_s$, and the emission angle with respect to the surface normal $\theta$, expressed as $\mu=\cos \theta$ (see \citealt{holai01} for details). For practical reasons, we specified the energy grid in terms of $\log(E/kT_{\rm eff})$ and the specific intensity grid in terms of $\log(I/T_{\rm eff}^3)$, because $E/kT_{\rm eff}$ and $I/T_{\rm eff}^3$ are relativistically invariant quantities.
The ranges of $\log T_{\rm eff}$ covered in our tables were $[5.1,6.8]$ for the hydrogen atmosphere and $[5.1,6.5]$ for the helium atmosphere. The range for $\log g_s$ was $[13.7,14.7]$, the range for $\mu$ was $[1\times10^{-6}, 1 -1 \times 10^{-6}]$ (plus three additional values near $\theta=0$ and $\theta=\pi/2$ to improve the grid resolution near, and avoid singularities at, these values of $\mu$), and the range for $\log(E/kT_{\rm eff})$ was $[-1.3,2.0]$. Based on tests, we found that spacings of the entries within these ranges of $\Delta(\log T_{\rm eff})=0.05$, $\Delta(\log g_s)=0.1$, $\Delta\mu=1/60$, and $\Delta\log(E/kT_{\rm eff})=0.0338$ produced interpolation errors that were negligible for our purposes ($\lesssim 0.1\%$). 
To obtain the specific intensity for a particular observed photon energy $E$ from a surface element with an effective temperature $T_{\rm eff}$, we took the corresponding value of $\log(E/kT_{\rm eff})$ and (1)~computed the surface gravity $g_s$ at the colatitude $\theta_c$ of the surface element in question, using the expression for the surface gravity given by \citet{2014ApJ...791...78A} for the assumed combination of NS mass, radius, and rotational frequency, (2)~determined the value of $\mu$ for which a photon from that point on the surface reaches the observer (for the given combination of surface element colatitude and rotational phase $\phi$; see Paper II for details), (3)~identified the entries in the look-up table that bracket the required values of $\log(E/kT_{\rm eff})$, $\log T_{\rm eff}$, $\log g_s$, and $\mu$, and then (4)~performed quadratic or quartic polynomial \citep{miller19}, or cubic polynomial \citep{riley19}, interpolation to obtain the corresponding value of $\log(I/T_{\rm eff}^3)$.

While the physics behind light element non-magnetic neutron star atmospheres is well understood, we caution that there may exist differences in the details of the numerical implementation, such as the discretization scheme, the depth considered for the atmosphere, the criterion for convergence, etc. With this in mind, we compared the output of NSX with the publicly available H atmosphere model McPHAC\footnote{Available for download at \url{https://github.com/McPHAC/}.} \citep{haakonsenetal12}.
In comparing NSX with McPHAC, we identified two major weaknesses of McPHAC which greatly limit its convergence to a consistent solution of the radiative transfer equation (RTE).  First, the angular grid and integration make use of an even number of Gauss-Legendre abscissas.  This leads to a lowest absolute value of the abscissa that differs significantly from zero, unless a very large number of grid points is used, and as a result of this non-zero value, McPHAC runs slowly.  This problem can be rectified by a trivial modification of McPHAC, so that it computes an odd number of abscissas for the angular grid and allows the lowest value to be set arbitrarily close to zero.  We find that such a modification yields a greater than ten-fold improvement in computational speed for the cases we consider.  Second, the temperature correction scheme used by McPHAC to converge to the solution of the RTE has several defects, as described in \citet{mihalas78}. We further note that in the publicly available version of McPHAC, the physical constants used are either outdated or are defined with low precision. In our comparisons against NSX, and the single hot spot test described in Section~\ref{sec:onespot}, we modified the values of these constants to their current and full precision values. In the two hot spot tests described in Section 3.2, we used the NSX nonmagnetic, fully ionized hydrogen-atmosphere model.

\subsection{Modeling the Hot Spots}
\label{sec:modeling-hot-spots}

In the parameter-estimation codes used by \citet{miller19} and \citet{riley19} to model the \textit{NICER} data, the X-ray emitting areas on the stellar surface were represented as a combination of single-temperature regions (see both papers for more in-depth descriptions of their modeling procedures, their choices of priors, and their treatment of degeneracies).

In \citet{riley19}, each single-temperature region is represented geometrically by a spherical cap with a size specified by an angular radius, with regions allowed to overlap in different ways. If a hot region is constituted by one such single-temperature region, it has the morphology of a hot spot. More complex hot regions can be constituted by two single-temperature regions that partially or wholly overlap, where one takes precedence. An order of precedence means that when evaluating the radiation intensity at a spacetime event (on the surface) that belongs to the intersection of two single-temperature regions, one of the regions supersedes the other, meaning one temperature is ignored at that spacetime event. Moreover, one of the single-temperature regions can have a temperature so low that the X-ray signal it generates is not computed because it would contribute negligibly in the \textit{NICER} waveband; however, if that (effectively) non-radiating region also takes precedence, the overall hot region is not a simple spherical cap. This implementation permits the representation of a variety of hot region geometries; for example, symmetric or asymmetric ring and crescent shapes,\footnote{A ring can be constructed if the single-temperature region that takes precedence has a low enough temperature so as not to contribute significantly to the observed emission and is itself a proper subset of the other single-temperature region. Similarly, a crescent is constituted by two such regions where the overlap is only partial. See \citet{riley19} for further details.} and two-temperature configurations where the hotter component is itself ring-like, a crescent, or simply a spherical cap. Finally, when two disjoint regions on the star are considered, these hot regions are not permitted to overlap (which would require extension of the order of precedence); this condition is used to define the prior support of hot region configurations. For the PSR~J0030$+$0451 analysis in \citet{riley19}, one of the X-ray emitting areas was found to be best represented by a hot crescent (a cooler spherical-cap component was shown not to be useful).

In their modeling of soft X-ray pulse profiles, \citet{miller19} start with the simplest possible description of the heated region(s) on the stellar surface and then, as necessary, consider more complicated descriptions, guided at each step by the differences between the best-fit simpler pulse  profile models and the pulsations observed using \textit{NICER}. Their algorithm allows circular and oval hot spots that can overlap in any way that improves the fit.  In the case of PSR~J0030$+$0415, they began by considering models with two and then three and four different uniform-temperature circular hot spots and were led to models with two and three different uniform-temperature oval spots. As the fitting proceeds, their algorithm allows the heated regions in the model to evolve toward a variety of different complex shapes, including multiple separate hot spots, multiple-temperature spots in which each spot has hotter and cooler regions, elongated or more complicated configurations produced by two or more adjacent or partially or completely overlapping circular or oval hot spots, as well as crescent-shaped hot regions produced by dark circular or oval spots that partially or completely overlap hot circular or hot oval spots, and so on.  When analyzing the PSR~J0030$+$0415 data, \citet{miller19} found no evidence for different temperatures within the same hot spot and no evidence for more than three heated regions on the stellar surface. They found that a model with three different, non-overlapping, uniform oval spots is preferred over any models in which the spots overlap, describes the observed energy-resolved pulse profile adequately, and gives a fit to the data that is slightly, but not significantly, better than the best-fit model with two non-overlapping, uniform oval spots.

In the parameter estimation codes used by \citet{miller19} and \citet{riley19}, the simpler spot models are subsets of the more elaborate spot models. An important aspect of the modeling approaches is that the complexity of the hot spot arrangement is increased, until a configuration that provides a satisfactory description of the data is reached. \citet{miller19} increased the complexity of the configuration still further to confirm that more complex spot patterns do not provide significantly better descriptions of the data. Despite the different approaches used by \citet{miller19} and \citet{riley19} in modeling the heated regions on the surface of PSR~J0030$+$0451, the marginalized constraints on $M$ and $R$ found using these different approaches are consistent with one another, with 68\% credible regions that have similar dimensions (spread/area) and largely overlap. This suggests that the mass and radius estimates obtained using these two approaches are relatively insensitive to the differences between them, which included different assumed priors on parameters (such as the stellar mass and radius) that are common to both models.

\subsection{Relativistic Ray-Tracing} \label{sec:ray-tracing}

To model the propagation of photons from the surface of a rapidly rotating MSP to Earth, we use the Oblate Schwarzschild (OS) approximation proposed by \citet{morsink07} and developed further by (\citealt{2014ApJ...791...78A}; see also \citealt{2020arXiv200805565S}). In this approximation, the spacetime in the vicinity of the star is assumed to be the Schwarzschild spacetime, but the rotation-induced oblateness of the stellar surface is explicitly taken into account using analytical approximations to describe the shape of the surface and the effective surface gravity as a function of colatitude. This model also incorporates all the relativistic effects that are important for the MSPs that are \textit{NICER} targets for determining neutron star masses and radii, including gravitational light bending, redshift, and lensing, and propagation time delays, Doppler boosts, and aberration. Comparing the predictions of this approximation with the results of exact numerical computations shows that it is accurate to within $\sim\,$0.1\% for spin frequencies $\lesssim\,$300~Hz (see \citealt{bogdanov19b}), which is more than adequate for our current parameter estimations. 

The details of this model, its implementation, and the verification tests of the codes we have used to analyze the \textit{NICER} pulse profile data are described in Paper~II \citep{bogdanov19b}. In brief, given the mass $M$ and equatorial radius $R_{\rm eq}$ of the NS and the locations, sizes, shapes, and temperature distributions of the X-ray emitting regions on the stellar surface, the observed flux is computed as a function of photon energy and the rotational phase of the star. The resulting set of model pulse profiles is then compared with the energy-resolved pulse profile observed using \textit{NICER}, to estimate $M$ and $R_{\rm eq}$.

In the mass and radius parameter estimation analysis presented in \citet{miller19}, the light-bending integral, lensing factor, and propagation time-delay integral are evaluated for a given dimensionless stellar radius ($R/M$) and photon emission angle $\theta$, and a particular combination of the star's rotational phase $\phi$ and the colatitude $\theta_c$ of the surface element being considered, by interpolating in a set of pre-computed lookup tables. These tables cover $R/M$ values in the range $[3.1,8.0]$, which includes the values of $R/M$ for all expected $M$ and $R$ combinations, and $\theta$ values in the range $[0,2.2]$~rad. Only photon trajectories that do not intersect the oblate stellar surface are considered. In \citet{riley19}, the deflection angle and propagation time delays at a set of colatitudes on every surface are computed on the fly and spline interpolation of these is then used, which also yields the lensing factor. The shape of the stellar surface is parameterized by $M$, $R_{\rm eq}$, $\Omega$. The arrays are bounded by the analytic most inward going ray and the radial ray, for every colatitude. Interpolation in pre-computed lookup tables dramatically reduces the computational cost of producing the vast number of energy-resolved pulse profiles that are needed for the sampling algorithms used for Bayesian inference.

\textit{Multiple imaging}.---For stars that are sufficiently compact, there can be ray paths from a given emitting region to the observer that proceed directly to the observer or encircle the star one or more times, causing the observer to see multiple images of the emitting region. In order to be sure that our Bayesian analyses could find such compact configurations if any of the \textit{NICER} MSPs are this compact, we cross-verified the ray paths and resulting pulse profiles for very compact stars that are given by the X-PSI\footnote{Available for download at \url{https://github.com/ThomasEdwardRiley/xpsi}. For X-PSI, internal cross-checks of the ``star-to-observer'' ray-tracing approach were conducted by validating against a light-curve integrator that discretizes a moderately distant image-plane (the so-called ``observer-to-star'' ray-tracing method).} package (referred to as the Amsterdam code in Paper II) and the Illinois-Maryland ray-tracing code.

\begin{figure}
\centering
\vspace{-60pt}
\includegraphics[width=0.55\textwidth]{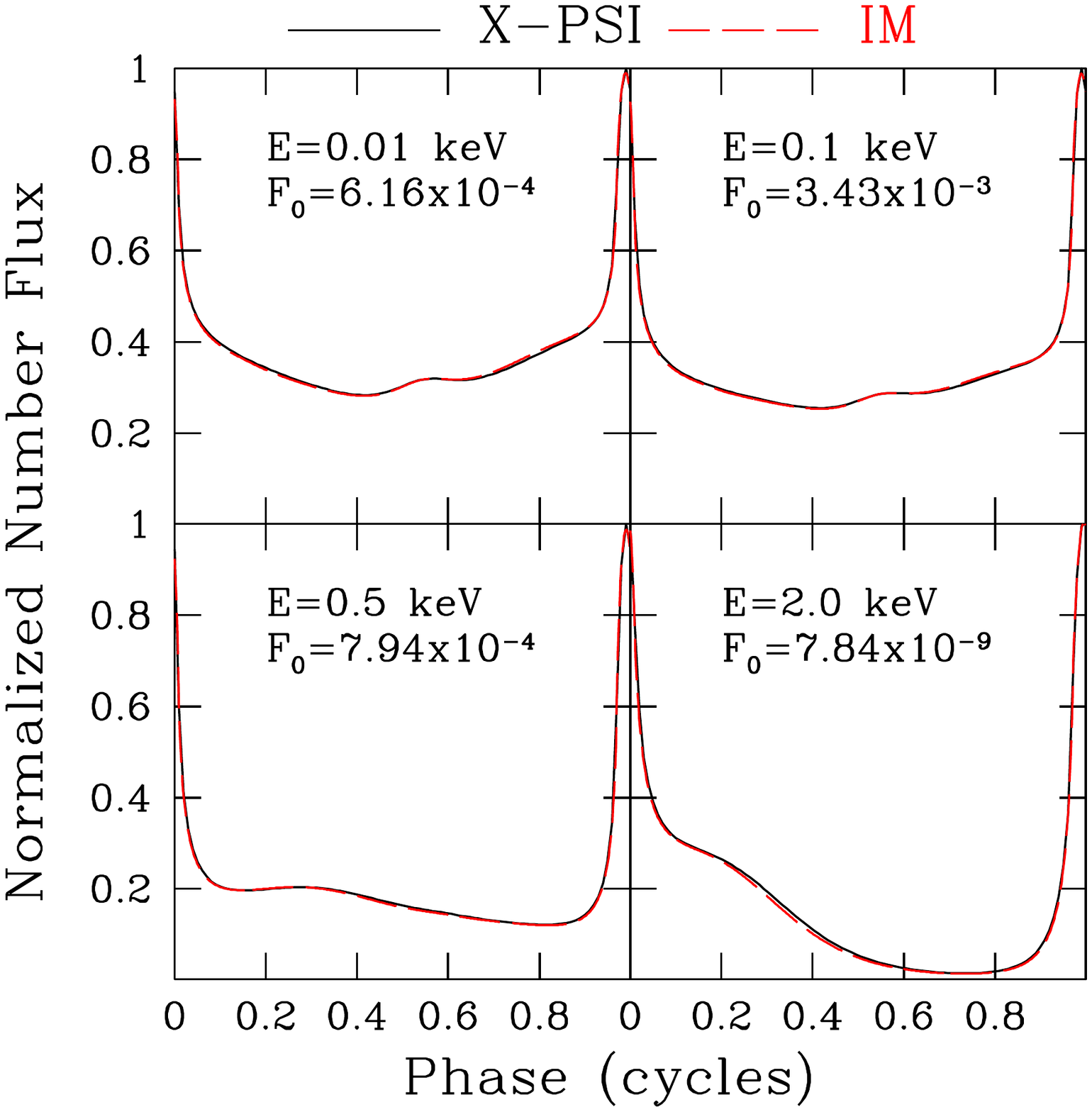}
\vspace{-90pt}
\caption{
Comparison of the pulse profiles at four different photon energies computed by the X-PSI (black) package and the Illinois-Maryland (red) ray-tracing code, for a highly compact ($R_{\rm eq}/M \approx 3.0$) star rotating at 346~Hz, with two non-identical, non-antipodal, uniform-temperature circular hot spots; see text for additional details.  Each panel is labeled by the observed energy and the maximum number flux $F_0$ (in units of ${\rm photons~cm}^{-2}~{\rm s}^{-1}~{\rm keV}^{-1}$) over all phases at that energy.  
}
\label{fig:compact-star-profiles}
\end{figure}

The most compact, theoretically stable, non-rotating stars that are possible when the matter in them has a nonnegative isotropic pressure have $c^2R/GM = 9/4$ (the ``Buchdahl limit''; see \citealt{1959PhRv..116.1027B}). We first verified our ray-tracing codes by computing the photon number flux as a function of energy that would be observed from a nonrotating star with $c^2R/GM = 3.0$, emitting blackbody radiation uniformly from its entire surface. Our codes agreed with the analytical result for this case to within a few parts in $10^5$. We then computed the pulse profiles produced by two stellar models with rotation frequencies $\nu_{\rm rot} = 346$~Hz, distance $1.0$~kpc, $M = 2.14\,M_\odot$, and $R_{\rm eq}=9.6$~km so that $c^2R_{\rm eq}/GM = 3.038$. One model assumed two identical, antipodal, uniform-temperature circular hot spots while the other assumed two non-identical, non-antipodal, uniform-temperature circular hot spots. In the non-antipodal comparison that we feature in Figure~\ref{fig:compact-star-profiles}, the observer inclination is $\theta_{\rm obs}=1.0$~radians; the first spot has a center at $\theta_{\rm c1}=\pi/2$ radians, an angular radius $\Delta\theta_1=0.2$~radians, and an effective temperature measured at the surface of $T_{\rm eff,1}=10^{6.1}$~K; and the second spot has a center at $\theta_{\rm c2}=\pi-1$~radians, an angular radius $\Delta\theta_2=0.1$~radians, and an effective temperature measured at the surface of $T_{\rm eff,2}=10^{6.3}$~K.  The longitude of the center of spot 2 is 0.55 cycles ahead of the longitude of the center of spot 1.  The pulse profiles given by the two ray-tracing codes agree to within 2\% or better in every pulse-phase bin and energy channel where the number flux is significant. Even at 2~keV, where the flux is more than five orders of magnitude smaller than the flux at 0.1~keV, the pulse profiles given by the two codes agree with one another to within 5\% or better at all pulse phases. This difference is a small fraction of the statistical uncertainties in the pulse profiles that have been observed to date using \textit{NICER}.

\subsection{Interstellar Absorption}
\label{sec:ism}

As X-ray photons propagate from the source to the observer, a fraction of them experience photoelectric absorption and scattering by the intervening interstellar medium (ISM). The observed spectral flux $f(E)$ is described by the expression $f(E)=\exp^{-\sigma_{\rm ISM}N_{\rm H}}f_0(E)$, where $f_0(E)$ is the emitted spectral flux, $\sigma_{\rm ISM}$ is the attenuation per unit column density of hydrogen, which depends on the composition and ionization state of the ISM, and $N_{\rm H}$ is the column density of hydrogen along the line of sight from the source to the observer. To obtain the attenuation of the X-ray flux between our target MSPs and the \textit{NICER} instrument for a given $N_{\rm H}$, we use the state-of-the-art \texttt{TBabs} model for $\sigma_{\rm ISM}$ and elemental abundances in the ISM \citep{2000ApJ...542..914W}. 

In computing models of the observed energy-resolved pulse profiles, \citet{miller19} and \citet{riley19} used a pre-computed look-up table that provides the attenuation at each \textit{NICER} energy channel for a particular, fiducial value of $N_{\rm H}$. Because the attenuation factor at every energy scales the same way with $N_{\rm H}$, \citet{miller19} simply scaled the attenuation factor at every energy to obtain the value appropriate for the value of $N_{\rm H}$ being considered at each step in the fitting process.

The energy dependence of $\sigma_{\rm ISM}$ alters the observed X-ray spectrum of an MSP, and therefore could in principle affect estimates of its mass and radius made using the \textit{NICER} data. However, the MSPs that are targets for $M$ and $R$ determinations using \textit{NICER} have exceptionally low values of $N_{\rm H}$ ($\sim 10^{20}$ cm$^{-2}$), so interstellar extinction has only a small effect on their spectra. Furthermore, the pulse-profile fitting method for estimating $M$ and $R$ uses the full energy-resolved profile, and is therefore relatively insensitive to changes in the observed spectrum produced by interstellar extinction, which are pulse-phase independent. Thus, the constraints on the masses and radii of these MSPs obtained using this method are insensitive to uncertainties in the chemical composition of the ISM along the line of sight \citep[see, e.g.,][]{2016ApJ...831..184B}, unlike constraints on the masses and radii of NSs in quiescent low-mass X-ray binaries (qLMXBs) that are based solely on their spectra and typically have to take into account high absorbing columns ($\gtrsim 10^{21}$ cm$^{-2}$) along the line of sight (see \citealt{2014MNRAS.444..443H}).

\subsection{Instrument Performance}
\label{instrument-performance}

In order to compare an energy-dependent model pulse profile with a pulse profile observed using \textit{NICER}, the performance of the telescope must be taken into account. The \textit{NICER} data consist of pulse-height events (not all of which are produced by X-ray photons from the source, or even by X-ray photons) in the $N$ energy channels of the detector. The spectrum registered by the instrument, which is reported in units of counts s$^{-1}$ channel$^{-1}$, is related to the photon spectrum $f(E)$ incident on the telescope, which is usually expressed in units of photons s$^{-1}$ keV$^{-1}$, by the convolution equation
\begin{equation}
    f_{\rm obs}(N)=\int f(E)R(N,E)\diff E  \;.
\label{eqn:convolution}
\end{equation}
Here $R(N,E)$ is the instrument response expressed as the probability that an incoming photon of energy $E$ will be detected in channel $N$. As is also the case for most other current X-ray instruments, the response of \textit{NICER}'s X-ray Timing Instrument (XTI) is described by an RMF and an ARF. The RMF describes the probability that a photon with a given energy that is incident on the instrument is counted in each of its output channels. The ARF describes the efficiency of the instrument in counting incident photons of a given energy, expressed as the effective area (in cm$^{2}$) of the instrument as a function of energy, and reflects the performance of the concentrator optics in focusing the incident X-rays as well as the quantum efficiency of the detector. 

In order to compare a model spectrum $f(E)$ with the spectrum $f_{\rm obs}(N)$ registered by the XTI, the model spectrum must be convolved with the instrument response, as described by Equation~(\ref{eqn:convolution}). We have verified this procedure by comparing the results given by the independently developed pulse-profile modeling codes described in Paper~II, to ensure that it is carried out correctly. We have also verified this procedure by comparing the results it gives for simple power-law and blackbody incident spectra with the results given by XSPEC \citep{arnaud96}, when the same instrument response files are used.

\subsection{Backgrounds}
\label{sec:backgrounds}

The final component needed to fit a pulse-profile model to the \textit{NICER} data is a model of the background, which we define as all counts accumulated by the \textit{NICER} XTI that do not come from the X-ray emitting hot spots on the neutron star surface. The models that best fit the  PSR~J0030$+$0415 pulse profile data collected by \textit{NICER} indicate that a significant fraction of the counts does not come from the hot spots \citep{miller19, riley19}, consistent with prior expectations of the source and background properties in XTI observations. These additional counts can be attributed to emission from gas in the immediate vicinity of the pulsar, to sky background (from unrelated neighboring sources and the diffuse X-ray background), and to non-cosmic photons and particles in \textit{NICER}'s local environment. Both \citet{miller19} and \citet{riley19} assumed that these non-spot background counts are not modulated at the rotation frequency of the pulsar, and the number of such counts in each energy bin was treated as a free parameter.\footnote{For details about how this background was modeled in \citet{miller19}, see Section~3.4 of that paper. For details about how it was modeled in \citet{riley19}, see Appendix~B of \citet{Riley19b}.} 

For future parameter estimation analyses using \textit{NICER} of rotation-powered MSPs, we are developing the capability to incorporate information about the \textit{NICER} non-source background emission estimated using the methods described in \citet{bogdanov19a}, as well as from any available archival X-ray observations of the same target with imaging telescopes such as \textit{XMM-Newton}, and \textit{Chandra}.

\subsection{Verification of Numerical Codes}
\label{sec:code-verification}

In our numerical computations of model pulse profiles, we used the most up-to-date values, and all significant figures, of the relevant physical and astrophysical constants (e.g., Planck's constant, the Boltzmann constant, the parsec, and $GM_{\odot}$) published by the Particle Data Group\footnote{See \url{http://pdg.lbl.gov/2017/reviews/rpp2017-rev-phys-constants.pdf} and \url{http://pdg.lbl.gov/2017/reviews/rpp2017-rev-astrophysical-constants.pdf}.}.

A critical requirement for producing accurate and reliable estimates of the masses and radii of the target MSPs is that the numerical codes used to compute model pulse profiles be free of errors and have sufficient numerical precision. We sought to assure this by completing a set of code comparison and pulse profile verification tests, using several independently developed codes to compute the model pulse profiles. As demonstrated in Paper~II \citep{bogdanov19b}, we obtained excellent agreement between the results given by these different codes for the same assumed input parameters, and between their results and a number of known analytical and semi-analytical results.

We made sure that the spacing between the entries in the lookup tables we used to compute the spectra and beaming patterns of the radiation from our model atmospheres and to perform ray-tracing in the vicinity of the neutron star models was fine enough that interpolation in these tables did not introduce numerical errors larger than the $\sim$0.1\% fractional error we established for the pulse profile comparisons in Paper~II and for the model pulse profiles we use to perform our Bayesian parameter estimation.

\subsection{Model Fitting and Parameter Estimation}
\label{sec:model-fitting}

By following the steps described above, we constructed energy-resolved models of the thermal X-ray pulse profiles produced by rotation-powered MSPs. We then used Bayesian techniques (see, e.g., \citealt{lo13,miller15,miller19}; and \citealt{riley19} for details) to estimate the best-fit values and credible intervals of all the parameters in a given pulse profile model, including the mass and radius of the pulsar, by fitting these models to the pulse profiles observed using \textit{NICER}. In the next section, we describe some of the tests we performed to validate the algorithms and codes we used in this procedure.

\section{Verification of Parameter Estimation Algorithms}
\label{sec:verification}

Bayesian parameter estimation is widely used in astrophysics, and there are a number of publicly available software packages that can be used to sample the space defined by the model parameters and construct the posterior distribution for each parameter or any combination of parameters
\citep[e.g.,][]{2013PASP..125..306F,2009MNRAS.398.1601F,2015MNRAS.450L..61H,2016arXiv160603757B,Betancourt2017}.
The inference analysis performed by \citet{riley19} on the PSR~J0030$+$0451 \textit{NICER} data used the MultiNest nested sampling algorithm (\citealt{2009MNRAS.398.1601F}) via the X-PSI package, while the analysis performed by \cite{miller19} used a hybrid approach in which MultiNest was used to perform an initial analysis, followed by more detailed sampling using PT-emcee, a parallel-tempered Markov chain Monte Carlo algorithm (\citealt{2013PASP..125..306F}). 
In early tests, the collaboration also considered using a basic Metropolis algorithm, an evolutionary algorithm \citep{2016ApJ...833..244S} coupled with a Metropolis algorithm, and PolyChord \citep{2015MNRAS.450L..61H}, but found them to be less well suited for the problem at hand. 

Given the posterior distribution, the credible region (or regions) that contain a specified total probability can be defined uniquely in any number of dimensions as the minimum-volume region containing that probability.  This is the approach used in the Illinois-Maryland code and in the X-PSI\footnote{X-PSI interfaces with the GetDist package \citep{Lewis2019} to calculate the minimum-area (or highest-density) credible regions in two dimensions.} package to compute the two-dimensional credible regions reported in \citet{miller19} and \citet{riley19}. In these codes, the minimum volume is computed by integrating over the posterior from higher to lower probability densities until the specified total probability is reached.  To compute the one-dimensional posteriors reported in \citet{miller19} and \citet{riley19}, both groups constructed the cumulative marginalized probability distribution for a given parameter and then reported probability intervals that are symmetric around the median. For example, the 90\% credibility interval ranges from the 5th percentile to the 95th percentile of the cumulative distribution.

As explained in Section~\ref{sec:modeling-hot-spots}, a realistic model of an MSP X-ray pulse profile produced by multiple hot spots necessarily requires a large number of free parameters, some of which are partially degenerate. Fitting such a model to \textit{NICER} pulse profile data and extracting parameter estimates is therefore both algorithmically and computationally challenging. Moreover, the priors for the model parameters need to be chosen carefully, so as not to produce specious posteriors\footnote{See \citet{miller19} and \citet{riley19} for a description of the priors used in the analyses of the \textit{NICER} data of PSR J0030+0451.}.  While the publicly available sampling and inference software packages have been thoroughly verified and used widely, they possess advantages and disadvantages depending on the particular problem in question \citep[see, e.g., Table 1 of][for a helpful summary]{2016arXiv160603757B}. For instance, with increasing number of parameters, certain algorithms may become very computationally inefficient, while for degenerate parameters and/or highly multi-modal distributions some algorithms may struggle with recovering all modes of the true posterior distribution. In light of this, it is crucial to verify whether the algorithms used for our parameter estimation with the \textit{NICER} data are implemented correctly and are capable of producing trustworthy inferences with the available tools and computational resources. To verify the ability of our algorithms and codes to overcome these problems, we conducted a number of tests. In them, our codes were used to analyze a series of progressively more complex synthetic \textit{NICER} pulse profile data sets that were generated using models of rotating neutron stars with various numbers of hot spots. The codes we used for analysis of \textit{NICER} data of PSR J0030+0451 in \citet{miller19}  and \citet{riley19} have been subjected to these tests and have been chosen based on their performance relative to other samplers.

In the first two subsections that follow, we describe in detail the results we obtained for two of these tests, which were carried out early in our development of parameter estimation procedures, prior to the analyses of the PSR J0030+0451 \textit{NICER} data set. In these tests, our parameter-estimation codes were used to analyze two different sets of synthetic \textit{NICER} pulse profile data, one produced by a model pulsar with a single, uniform-temperature, circular hot spot, and the other, by a model pulsar with two different, uniform-temperature, circular hot spots. As we show below, the analysis procedures used by both \citet{miller19} and \citet{riley19}\footnote{This test used a development version of the X-PSI package that preceded \texttt{v0.1}.} to analyze these two synthetic pulse profiles produce estimates of the values of the model parameters and their credible regions that are consistent with the parameter values used to generate the synthetic data.

In the third subsection that follows, we describe two additional checks that \citet{miller19} performed to test the performance of their inference code by using it to analyze synthetic pulse profiles.  The first of these additional checks was performed by analyzing a third synthetic pulse profile that was specially constructed to mimic the pulse profile of PSR J0030+0451 that was observed using \textit{NICER}. This synthetic profile was generated using a model pulse profile produced by two non-overlapping, oval hot spots and was fit using a model with two possibly overlapping oval hot spots. As we discuss below, this fit gave estimated values and credible regions for all of the 15 parameters in the model that are consistent with the values that were used to generate the synthetic profile (see Section~4.2 of \citealt{miller19}).

The second additional check was made by performing posterior-predictive assessments of whether the models with two and three oval hot spots used by \citet{miller19} are able to accurately describe the PSR~J0030$+$0451 pulse profile and unmodulated background observed using \textit{NICER}. As we detail below, both models pass this test.

We note that due to the complexity of the models we are using and the consequent high computational cost of analyzing the \textit{NICER} data, at present it is not feasible to conduct a large-scale simulation-based calibration \citep[see][]{Berry2015,Talts2018,Riley19b} of our inference procedures, to demonstrate that our algorithms are capable of thoroughly exploring the vast parameter space and obtaining sufficient credible region coverage. The tests and cross-comparisons we have conducted do demonstrate that (1)~the model components discussed in Section~\ref{sec:model} have been correctly implemented in our data-fitting algorithms, and (2)~our inference codes yield parameter estimates and credible regions that are consistent with the values assumed in generating several different sets of synthetic pulse profile data.

\subsection{A Neutron Star with a Single, Uniform-Temperature, Circular Hot Spot}
\label{sec:onespot}

In the first test of our algorithms, we analyzed a synthetic \textit{NICER} pulse profile produced by a spinning neutron star with a single, uniform-temperature, circular hot spot. In constructing this synthetic pulse profile, we used the following procedure:\footnote{The assumed parameter values for this test are largely the same as those assumed in Test~OS1b presented in Paper~II. The synthetic data set used in this analysis, the tabulated atmosphere model, and the \textit{NICER} instrument response files are available for download from Zenodo.}

\begin{enumerate}

\item
We chose a gravitational mass $M=1.4$\,$M_{\odot}$ and a circumferential equatorial radius $R_{\rm eq}=12.0~{\rm km}$ (thus $c^2R_{\rm eq}/GM=5.8047$).

\item
We assumed a stellar rotational frequency $\nu_{\rm rot} = 600$~Hz, as seen by a distant, static observer. The shape of the stellar surface and the effective gravity at the stellar surface as a function of rotational colatitude were computed using the \citet{2014ApJ...791...78A} model.

\item
We chose a distance to the neutron star of $D=200$\,pc.

\item
We assumed that the observer is situated in the plane defined by the star's rotational equator (observer colatitude $\theta_{\rm obs}=90^{\circ}$).

\item
We used the McPHAC nonmagnetic, fully ionized hydrogen-atmosphere model \citep{haakonsenetal12} As noted previously, while we have identified issues with the McPHAC code, for the purposes of this comparison of  parameter-estimation codes they are unimportant as both codes made use of the exact same atmosphere model lookup tables.

\item
We considered a circular, uniform-temperature spot centered on the star's rotational equator (colatitude of the spot center $\theta_c=90^{\circ}$) with an angular radius $\Delta\theta=1.0$~rad and an effective temperature $kT_{\rm eff}=0.231139$~keV, as measured in the surface comoving frame.

\item
For simplicity, in this simulation we neglected absorption by the ISM (i.e., we set $N_{\rm H}=0$~cm$^{-2}$).

\item
We considered only the counts in \textit{NICER} energy channels 0 through 299 inclusive (corresponding approximately to the energy band 0.1 to 3.09~keV).

\item
For each of these 300 energy channels, we computed a high-accuracy profile of the pulse produced by the hot spot at $32$ phase points (rather than integrating high-accuracy pulse profiles computed at closely-spaced phases over each of the 32 phase bins).

\item
We folded these high-accuracy synthetic pulse profiles through the pre-launch \textit{NICER} response matrix and then normalized this synthetic ``observed'' pulse profile so that the total expected number of counts in channels 0 through 299 (i.e., the expected number of counts from the hot spot) was $10^6$. This set the total observation time. This synthetic ``observed'' pulse profile was then sampled, to produce a set of synthetic pulse profile data.

\item
We assumed a phase-independent (unpulsed) background, with a power-law photon number spectrum ${\rm d}N/{\rm d}E \propto E^{-2}$.
Next, we folded this assumed energy-dependent but phase-independent background through the pre-launch \textit{NICER} response matrix, to produce the synthetic ``observed'' background. We then normalized this background so that the total expected number of background counts in channels 0 through 299 was $10^6$. (The expected total number of counts from the spot and from the unpulsed background was therefore $10^6+10^6 = 2 \times 10^6$.) Finally, the normalized ``observed'' background was sampled to produce the synthetic background data.

\end{enumerate}
We analyzed the synthetic pulse profile data (which included the synthetic background data) by fitting ``practical'' model pulse profiles\footnote{In principle, the codes used for these analyses can compute model pulse profiles to high precision using the OS approximation, but at increased computational cost. Computing these profiles dominates the time required to evaluate the likelihood. Consequently, it is necessary to strike a balance between precision and speed. Thus, when analyzing these synthetic pulse profile data and the pulse profile data collected using \textit{NICER}, we used so-called ``practical'' model profiles that have a fractional precision of only $\sim\,$0.1\%, but can be generated in $\sim\,1$~s per processor core (see Paper~II for more details).} to them, using the parameter-estimation algorithms developed by \citet{miller19} and \citet{riley19}. Both the synthetic pulse profile data and the model pulse profiles were binned in 32~phase bins\footnote{From previous comparisons we find that integrating over a phase bin or calculating the counts (after folding) per unit phase at the center of a phase bin and then multiplying by the width of the phase bin give answers that differ by an amount that is small compared with the uncertainty in the pulse profile data. Thus, either approach may be used, but it is helpful to specify which is assumed.}, which is a sufficient number of bins to accurately describe the smoothly varying pulse profile produced by the single-spot pulsar model considered in this test.
The independent model profiles that were fit to the synthetic profile data using the X-PSI package and the Illinois-Maryland code had nine free parameters. Eight of these parameters occur in both models, namely, the star's mass $M$ and equatorial radius $R_{\rm eq}$, the colatitude of the center of the hot spot $\theta_c$, the angular radius $\Delta\theta$ and effective temperature $T_{\rm eff}$ of the hot spot, the angle $\theta_{\rm obs}$ between the rotation axis of the star and the direction to the observer, the distance $D$ to the observer, and the hydrogen column density $N_{\rm H}$ of the ISM along the line of sight to the observer. In addition, the X-PSI analysis included as a free parameter the azimuthal phase of the hot spot, whereas the Illinois-Maryland analysis marginalized over the overall phase of the pulse, making a total of nine individual free parameters in both analyses. Both model pulse profiles also included an arbitrary, phase-independent additive background parameter in each energy channel.

\begin{figure}
\centering
\includegraphics[width=0.47\textwidth]{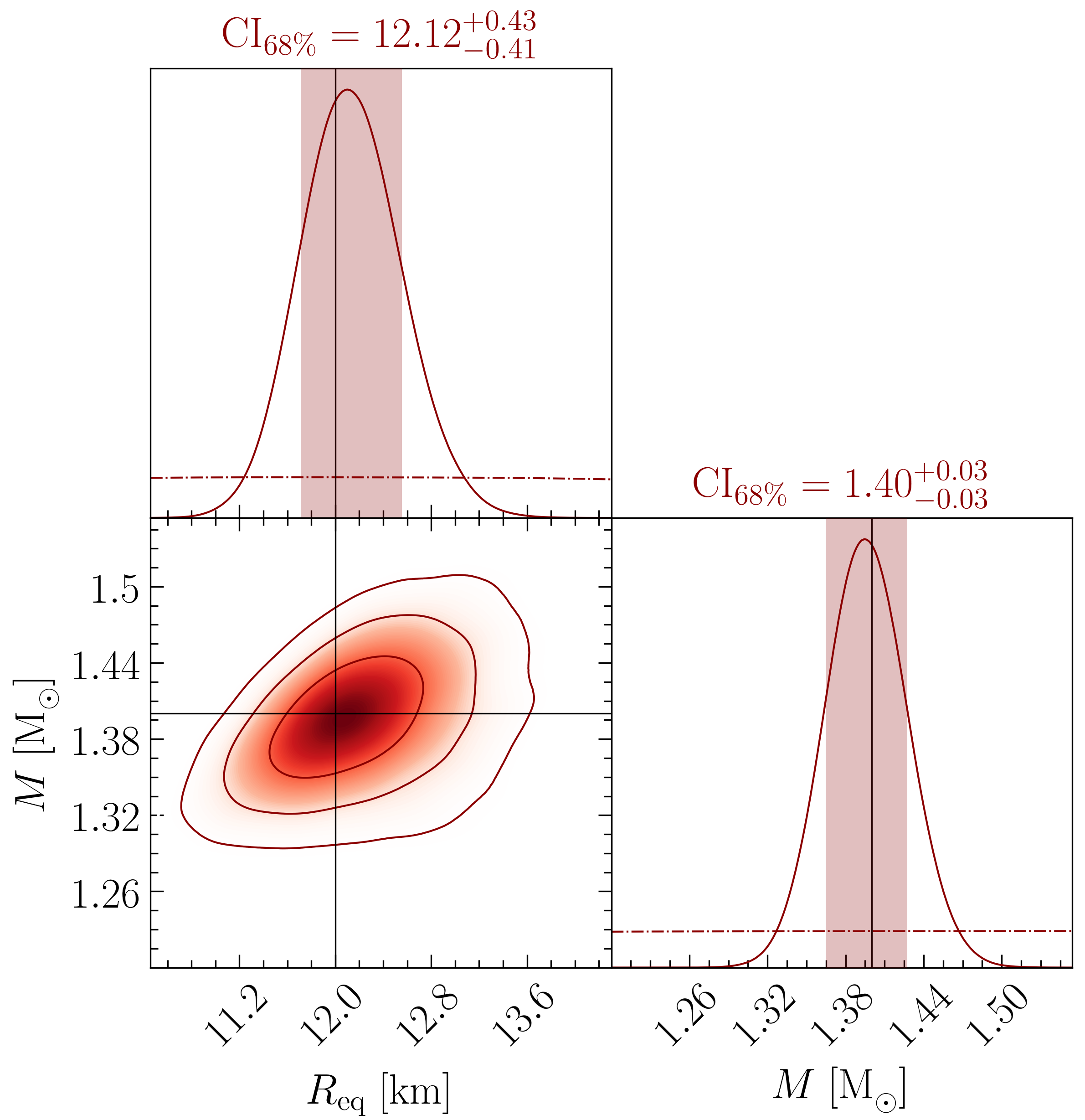}
\includegraphics[clip,trim = 1cm 1cm 2cm 0cm,width=0.52\textwidth]{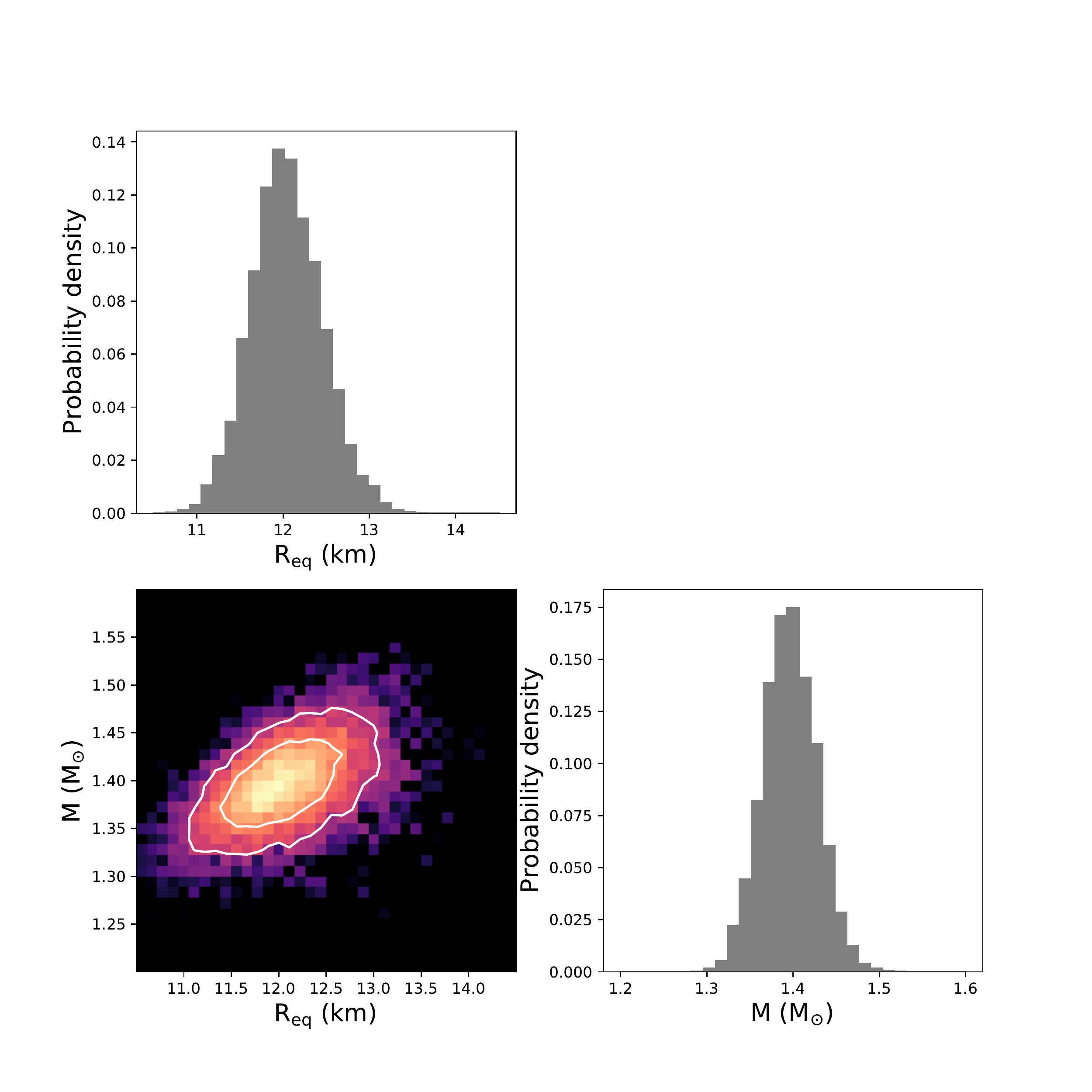}
\caption{
Posterior distributions in the $M$--$R_{\rm eq}$ plane obtained when the same synthetic \textit{NICER} pulse profile produced by the stellar model with a single hot spot described in Section~\ref{sec:onespot} is analyzed independently using the X-PSI package~(left) and the Illinois-Maryland code~(right).
Left: The red contours in the $M$--$R_{\rm eq}$ panel indicate the two-dimensional 1$\sigma$, 2$\sigma$, and 3$\sigma$ credible regions, while the pink bands show the 1$\sigma$ ranges of the one-dimensional posterior distributions. The black lines show the values of $M$ and $R_{\rm eq}$ that were assumed in generating the synthetic data set. The dashed-dotted lines in the one-dimensional plots show the prior and the vertical lines show the parameter values used to generate the synthetic data. 
Right: The $M$--$R_{\rm eq}$ panel shows the joint posterior probability distribution, colored by credibility interval; the white contours enclose the 1$\sigma$ and 2$\sigma$ credible regions. The grey histograms show the one-dimensional posterior probability density distributions for $R_{\rm eq}$ and $M$. The 1$\sigma$ credible region for $R_{\rm eq}$ is $12.03^{+0.42}_{-0.39}$, while the 1$\sigma$ credible region for $M$ is $1.40 \pm 0.03$. These two independent analyses produce results that are statistically indistinguishable and consistent with the $M$ and $R_{\rm eq}$ values assumed in generating the synthetic data. The residuals computed by comparing the best-fit pulse profile models with the synthetic pulse-profile data are acceptable.
}
\label{fig:one_spot_mr}
\end{figure}

The resulting joint posterior distributions of $R_{\rm eq}$ and $M$ given by the Illinois-Maryland code and the X-PSI package, and the corresponding one-dimensional posterior probability distributions for $R_{\rm eq}$ and $M$, are shown in Figure~\ref{fig:one_spot_mr}. This figure shows that (1)~these two independent analyses of the same synthetic data set produced results that are statistically indistinguishable and (2)~the credible regions given by both algorithms are consistent with the $M$ and $R_{\rm eq}$ values that were assumed when generating the synthetic data, in that the injected values of $M$ and $R_{\rm eq}$ belong to a posterior mode. Moreover, the residuals computed by comparing the best-fit pulse profile models to the pulse-profile data do not show any statistically significant discrepancies.

\subsection{A Neutron Star with Two Different, Uniform-Temperature, Circular Hot Spots}
\label{sec:twospot}

As a second test of our analysis algorithms, we analyzed a synthetic \textit{NICER} pulse profile produced by a spinning neutron star with two different, uniform-temperature, circular hot spots. The properties and locations of these two spots were chosen to produce a synthetic \textit{NICER} pulse profile that qualitatively resembles the observed profile of PSR~J0437$-$4715 (see, e.g., Figure~2 of Paper~I). In constructing this synthetic pulse profile, we used the following procedure:

\begin{enumerate}

\item
We chose a gravitational mass $M=1.44$\,M$_{\odot}$ and a circumferential equatorial radius $R_{\rm eq}=13.0$~km (thus $c^2R_{\rm eq}/GM=6.1138$).

\item
We assumed a stellar rotational frequency $\nu_{\rm rot} = 173.6$ Hz, as seen by a distant, static observer. The shape of the stellar surface and the effective gravity at the stellar surface as a function of colatitude were again computed using the \citet{2014ApJ...791...78A} model.

\item
We chose a distance to the neutron star of $D=156.3$\,pc.

\item
We assumed that the inclination of the observer's sightline relative to the stellar spin axis is $\theta_{\rm obs}=0.733$~rad.

\item
We used the NSX nonmagnetic, fully ionized hydrogen-atmosphere model  \citep{holai01, hoheinke09}.
 
\item 
We considered a circular, uniform-temperature, primary spot centered at colatitude $\theta_{c1}=0.6283$~rad and azimuth $\phi=0$~rad, where the zero of azimuth is defined by the plane containing the observer, with an angular radius $\Delta\theta_{1}=0.01$~rad and an effective temperature $kT_{\rm eff,1}=0.231139$~keV, as measured in the surface comoving frame.

\item
We considered a circular, uniform-temperature, secondary spot centered at colatitude $\theta_{c2} = 2.077$~rad and azimuth $\phi_2 = 3.5343~{\rm rad} = 0.5625~{\rm cycles}$, with angular radius $\Delta\theta_2 = 0.33$~rad and an effective temperature $kT_{\rm eff,2} = 0.0577846$~keV, as measured in the surface comoving frame.

\item
We included absorption by the ISM, assuming $N_{\rm H}=2\times 10^{20}$ cm$^{-2}$ and using the elemental abundances given by \citet{2000ApJ...542..914W}.

\item
We assumed a total simulated observation time of $10^6$ seconds.

\item
We again assumed a pulse-phase independent background with the power-law spectrum $dN/dE \propto E^{-2}$, prior to folding through the \textit{NICER} pre-launch instrument response matrix, normalized to a total of $10^6$ expected counts in energy channels 25 through 299 inclusive after folding.
\end{enumerate}

When generating the synthetic pulse profile data analyzed in this test, and when computing the model pulse profiles used to fit these synthetic data, we used the following ancillary information: (i)~Look-up tables for the spectra and beaming patterns produced by the NSX nonmagnetic, fully ionized hydrogen-atmosphere model \citep{holai01, hoheinke09}, used to compute the properties of the radiation from the hot spots; (ii)~the look-up table described in Section~\ref{sec:ism} for the TBabs model, used to calculate the ISM absorption; and (iii)~the publicly available, post-launch (v1.02) \textit{NICER} RMF and the ARF, corrected for the off-axis pointing used when observing PSR~J0437$-$4715\footnote{As discussed in Paper~I, due to the presence of a bright AGN near the position of PSR~J0437$-$4715, a 1.47$'$ pointing offset was used when observing this pulsar using \textit{NICER}, to reduce contamination of the pulsar data by photons from the AGN.}, used to compute the pulse profile that would be observed using \textit{NICER}.
We again binned the synthetic pulse profile data and the model pulse profiles in 32~phase bins,  which is a sufficient number of bins to accurately describe the smoothly varying rotation-induced flux modulation  produced by the two-spot pulsar model considered in this test.

Due to the increased complexity of the two-spot pulse profile model relative to the single-spot model, we performed several preliminary checks on the two-spot model before proceeding to analyze the synthetic two-spot data. In these checks, we used the pulse profile modeling codes described in Paper~II to independently produce, and then cross-compare, the six intermediate model pulse profiles we now list. The first four preliminary profiles were lists of the number of photons\,cm$^{-2}$\,s$^{-1}$ in each of the $32 \times 700$ pulse-phase--energy-channel bins that describe the profile; the final two profiles were lists of the number of counts expected in each \textit{NICER} pulse-phase--energy-channel bins. These preliminary profiles were:
\begin{enumerate}
\item  The individual, unsampled pulse profiles produced separately by the first and the second hot spots, assuming no background and $N_{\rm H} = 0$. 

\item  The unsampled, combined pulse profile produced by both spots, again assuming no background and $N_{\rm H} = 0$.
 
\item  The unsampled, combined pulse profile produced by both spots, with ISM absorption included, assuming $N_{\rm H}=2.0 \times 10^{20}$ cm$^{-2}$.

\item  The unsampled, combined pulse profile produced by both spots, with ISM absorption included, assuming $N_{\rm H}=2.0 \times 10^{20}$ cm$^{-2}$, and including the power-law background. The ISM absorption was not applied to the background.

\item The unsampled, combined pulse profile produced by both spots after absorption by the ISM and the assumed background, after folding through the \textit{NICER} off-axis detector response. The result is a list of the number of counts in each of the 32 phase bins, for \textit{NICER} energy channels 25 through 299 inclusive.

\item  Finally, the realistic synthetic data set, consisting of the combined pulse profile produced by both spots after absorption by the ISM and the assumed background, after folding through the \textit{NICER} off-axis detector response and sampling. The result is again a list of the number of counts in each of the 32 phase bins, for \textit{NICER} energy channels 25 through 299 inclusive, but now sampled. These are the data that were analyzed for this parameter estimation exercise.
\end{enumerate}
For the four preliminary pulse profiles that were not folded through the detector response matrix, the photon flux in each energy bin was determined by integrating over each bin, rather than by computing the photon flux at the center of each bin. Each bin had a width of 0.005~keV; the low-energy edges of the bins were chosen to be 0.1~keV, 0.105~keV, 0.11~keV, ..., 3.590~keV, and 3.595~keV. The preliminary and final synthetic pulse profiles, expressed as phase-channel data,  the instrument response matrices, and tables of the extinction as a function of $N_{\rm H}$ are available on Zenodo.

\begin{figure}
\includegraphics[width=0.47\textwidth]{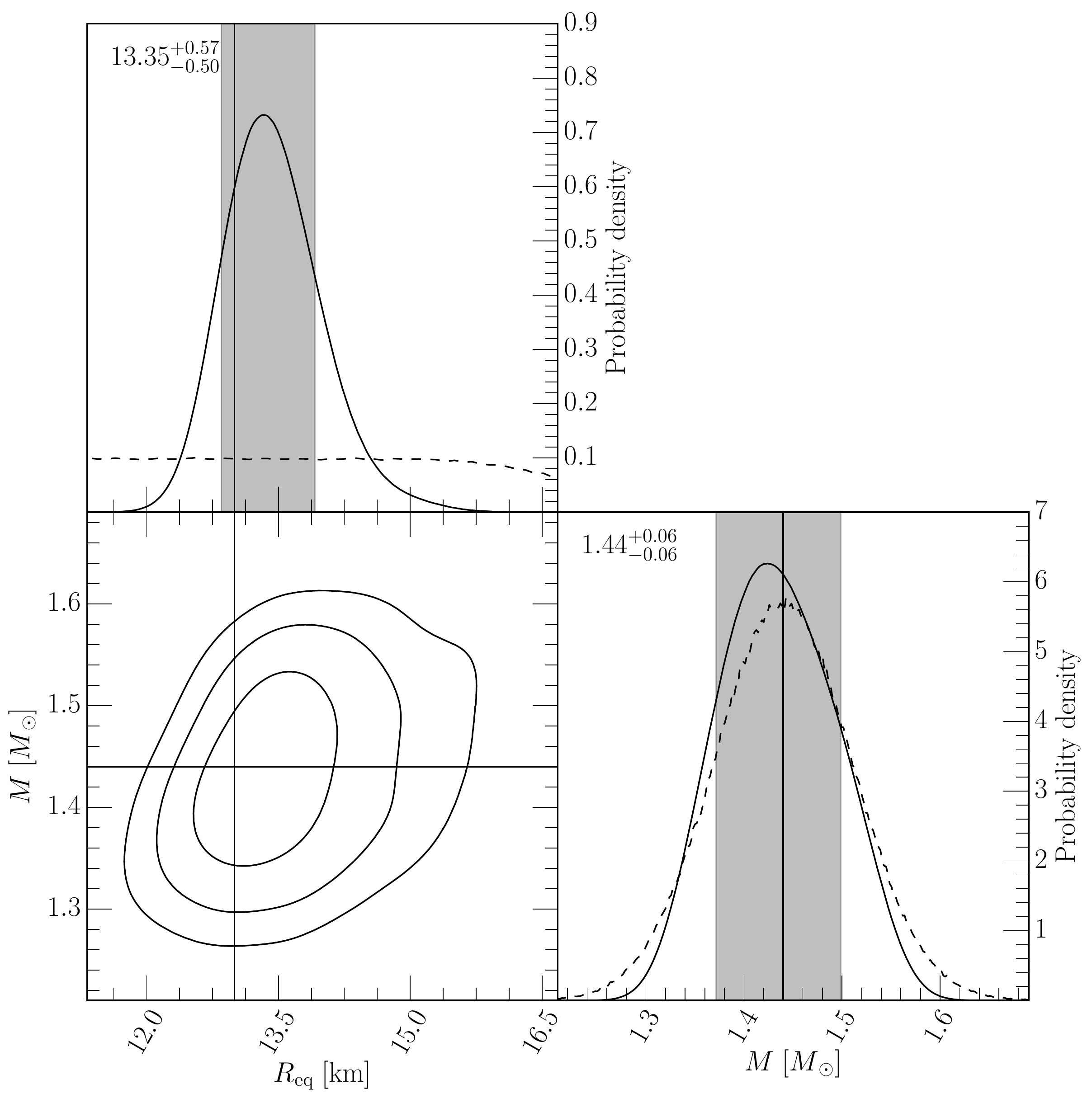}
\includegraphics[clip,trim = 1cm 1cm 2cm 0cm,width=0.52\textwidth]{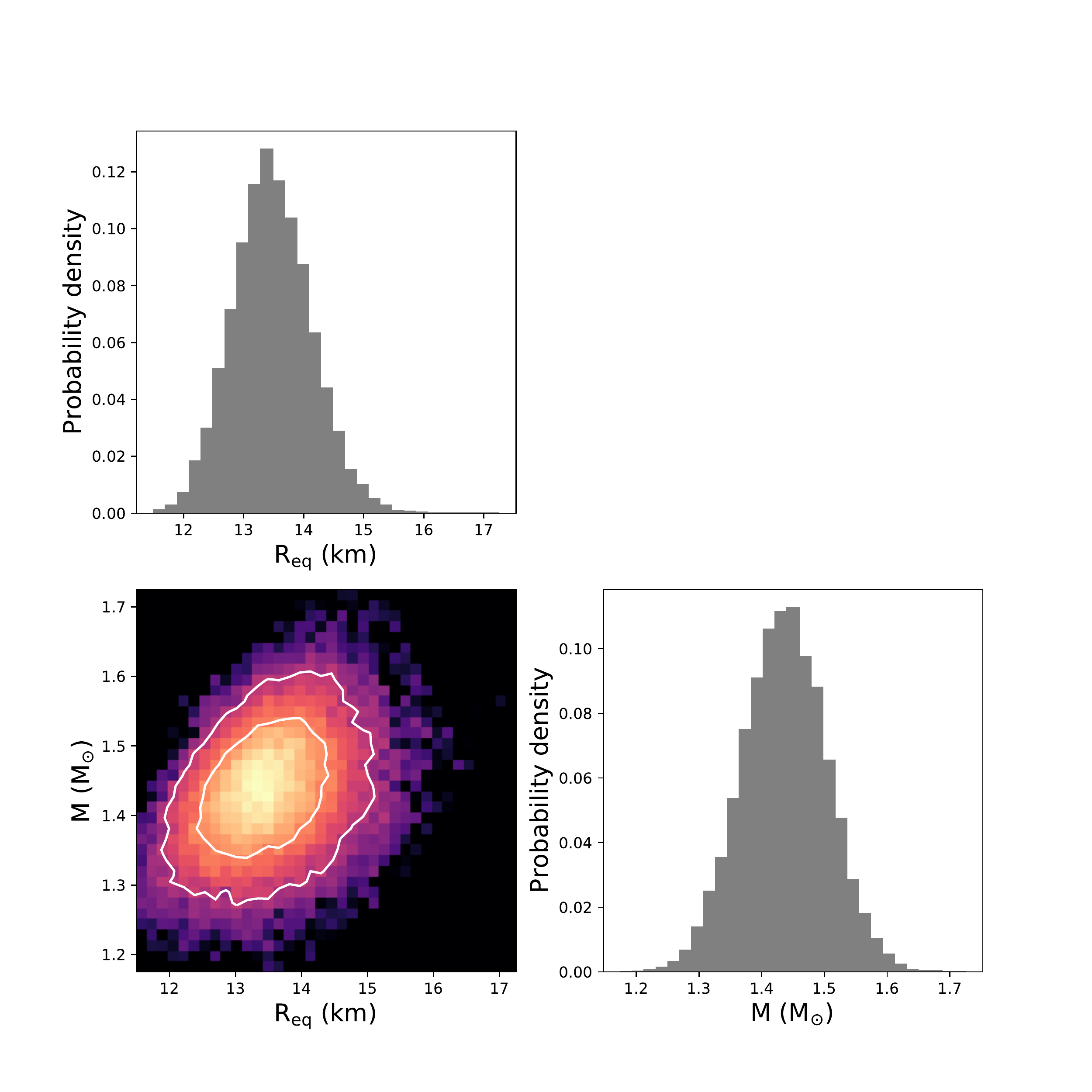}
\caption{
Posterior distributions in the $M$--$R_{\rm eq}$ plane obtained when the same synthetic \textit{NICER} pulse profile produced by the stellar model with two hot spots described in Section~\ref{sec:twospot} is analyzed using the X-PSI package~(left) and the Illinois-Maryland code~(right).
Left: The black contours in the $M$--$R_{\rm eq}$ panel indicate the two-dimensional 1$\sigma$, 2$\sigma$, and 3$\sigma$ credible intervals, while the grey bands show the 1$\sigma$ ranges of the one-dimensional posterior distributions. The dashed lines in the one-dimensional plots show the priors and the vertical lines show the parameter values used to generate the synthetic data. 
Right: The $M$--$R_{\rm eq}$ panel shows the joint posterior probability distribution, colored by credibility interval; the white contours enclose the 1$\sigma$ and 2$\sigma$ credible regions. 
The grey histograms show the one-dimensional posterior probability density distributions for $R$ and $M$. The 1$\sigma$ credible region for $R_{\rm eq}$ is $13.45^{+0.66}_{-0.62}$, while the 1$\sigma$ credible region for $M$ is $1.44^{+0.06}_{-0.07}$.
Again, these two independent analyses produce results that are statistically indistinguishable and consistent with the $M$ and $R_{\rm eq}$ values assumed in generating the synthetic data. The residuals computed by comparing the best-fit pulse profile models with the synthetic pulse-profile data are acceptable.
}
\label{fig:two_spot_mr}
\end{figure}

We again analyzed the realistic synthetic pulse profile data by fitting ``practical'' model pulse profiles to these data, using the parameter-estimation codes developed by \citet{miller19} and \citet{riley19}. The model profiles used by both groups have thirteen free parameters. Twelve are present in both models: the star's mass $M$ and equatorial radius $R_{\rm eq}$, the colatitude of the center of the primary spot $\theta_{c1}$, the angular radius $\Delta\theta_1$ and effective temperature $T_{\rm eff,1}$ of the primary spot, the colatitude of the center of the secondary spot $\theta_{c2}$, the azimuthal phase $\phi_2$, angular radius $\Delta\theta_2$, and effective temperature $T_{\rm eff,2}$ of the secondary spot, the angle $\theta_{\rm obs}$ between the rotation axis of the star and the direction to the observer, the distance $D$ to the observer, and the hydrogen column density $N_{\rm H}$ of the ISM along the line of sight to the observer. In addition, the X-PSI analysis included the azimuthal phase $\phi_1$ of the primary spot, whereas the Illinois-Maryland analysis marginalized over the overall phase of the pulse profile (an equivalent approach that makes the fitting procedure more efficient), making a total of thirteen individual free parameters in both analyses. Both model pulse profiles also included an arbitrary, phase-independent background in each energy channel.

The resulting joint posterior distributions of $R_{\rm eq}$ and $M$ given by the Illinois-Maryland code and the X-PSI package, and the corresponding one-dimensional posterior probability distributions for $R_{\rm eq}$ and $M$, are shown in Figure~\ref{fig:two_spot_mr}. This figure shows that (1)~these two independent analyses produced results that are statistically indistinguishable and (2)~the $M$ and $R_{\rm eq}$ values that were assumed when generating the synthetic data belong to a posterior mode for both algorithms. Moreover, comparing the best-fit pulse profile models to the pulse-profile data shows no statistically significant discrepancies in the residuals.
This finding offers assurance that the posterior computations are well-suited for the analyses of actual \textit{NICER} data.

\subsection {Analyses of Synthetic Pulse Profiles that Mimic the \textit{NICER} Data on PSR~J0030$+$0451}
\label{sec:analyses-J0030-synthetic-data}

The analyses of the PSR J0030+0451 \textit{NICER} data set by \citet{miller19} and \citet{riley19} revealed greater complexity in the hot spot shapes and arrangement on the stellar surface compared to the early synthetic data analyses described above. This raises the question of whether the parameter estimation procedures of \citet{miller19} and \citet{riley19} give correct posterior probability distributions for the range of pulse profile models applied to the NICER data on PSR J0030+0451. To address this,
in addition to testing their pulse profile modeling and inference procedures by analyzing the two synthetic pulse profiles produced by one and two circular spots as just described, \citet{miller19} also tested them by analyzing a third synthetic pulse profile that was specifically constructed to mimic the pulse profile of PSR~J0030$+$0451 observed using \textit{NICER}. We summarize this analysis and its results in the first paragraphs of this subsection. We then describe posterior predictive assessments of the fits of the two pulse profile models used by \citet{miller19} to the \textit{NICER} data on PSR~J0030$+$0451. Unlike the previous analyses of synthetic pulse profiles, these assessments include both the PSR~J0030$+$0451 pulse profile and the specific unmodulated thermal-like background observed using \textit{NICER}.

\textit{Analyses of synthetic pulse profile data generated using a model with two oval hot spots}.---In addition to their analyses of the two synthetic pulse profiles produced by one and two circular spots just described, \citet{miller19} tested their inference procedure further by analyzing a third synthetic pulse profile, specifically constructed to mimic the pulse profile of PSR~J0030$+$0451 observed using \textit{NICER}. This synthetic data was produced using a pulse profile model with two different, non-overlapping, uniform oval hot spots, with the model's 14 parameters chosen to mimic the pulse profile of PSR~J0030$+$0451, and assuming the \textit{NICER} instrument's response. This synthetic data was then fit using a 14-parameter model pulse profile that assumed two possibly different and overlapping uniform oval hot spots. The resulting fit gave estimated values and credible regions for all 14 of the parameters in the model that are consistent with the values of these parameters that were used to generate the synthetic profile (see also Section 4.2 in \citealt{miller19}). 

Specifically, \citet{miller19} found that the values of~12 of the~14 parameters assumed when generating the synthetic pulse profile are within the $\pm 1\,\sigma$ credible intervals derived from their posterior probability distributions and that the assumed values of the other two parameters are within the corresponding $\pm 2\,\sigma$ credible intervals (see Table~5 in \citealt{miller19}). Thus, the 1-D posterior probability distributions that were obtained by analyzing the synthetic data are all consistent with the values of the model parameters that were assumed when the synthetic pulse profile data was produced.
\citet{miller19} also found that the joint probability density distribution for $M$ and $R_{\rm eq}$ obtained by fitting this model to the synthetic \textit{NICER} pulse profile data are consistent with the values of the stellar mass and equatorial radius assumed when the synthetic data was produced (see Figure~3 in \citealt{miller19}).

\textit{Posterior predictive assessments of the fits of pulse profile models to the \textit{NICER} data on PSR~J0030$+$0451}.---In addition to the verification provided by the parameter-estimation analyses of the three synthetic \textit{NICER} pulse profile data sets that we discussed previously, \citet{miller19} provided information about the best fits of their models with two and three oval hot spots to the \textit{NICER} data on PSR~J0030$+$0451 that allows us to effectively perform posterior-predictive assessments of the ability of their models, which had (a)~the 205-Hz spin frequency of PSR~J0030$+$0451, (b)~two or three oval hot spots, and (c)~an arbitrary unmodulated background, to accurately describe the PSR~J0030$+$0451 pulse profile and the specific unmodulated thermal-like background observed using \textit{NICER}.

A posterior-predictive assessment of a fit to data is the Bayesian version of a goodness of fit test (see, e.g., Gelman, Meng, and Stern 1996, Statistical Sinica, 6, 733). It tests whether the log likelihood of the \textit{actual} data, given the model and a set of assumed values of the parameters in the model, is or is not far out on the low tail of the distribution of log likelihood values computed using a set of synthetic data; if it is, then according to this test the model is not a good fit to the data. 

In applying this test to the fits of their models to the \textit{NICER} data on PSR~J0030$+$0451 obtained by \citet{miller19}, we first note that the value of the phase-channel chi-squared reported by \citet{miller19} for the best fit of their pulse profile model with two oval spots to the PSR~J0030$+$0451 data had a probability of 0.104 if this model were correct, while the best fit of their model with three oval spots had a probability of 0.120 if that model were correct. This means that the actual PSR~J0030$+$0451 data collected by \textit{NICER} could plausibly be drawn in a Poisson realization of their best-fitting model with two oval spots, or from a Poisson realization of their best-fitting model with three oval spots, and with the unmodulated, backgrounds inferred from their fits.  

Hence, for a specific set of the values of the parameters in either of their models, one can compute the log likelihood of the data given the model for that particular set of parameter values. Using the same model and the same set of parameter values, one can also generate synthetic data by Poisson-sampling the expected number of counts in each phase-channel bin given by the model. One can then compute the log likelihood for the synthetic data given the model with those parameter values. Indeed, one can do this many times and thus obtain a distribution of the log likelihoods one would expect if the model were correct and the parameter values were the values assumed.

The information needed to perform this analysis was provided in \citet{miller19}: one simply compares the $\pm 1\sigma$ and $\pm 2\sigma$ credible intervals for each of the parameters in their two pulse profile models with the best-fit values of these parameters, which are listed in their Tables~7 and~8. The result is that 13 of the 15 parameters in the model with two oval hot spots fall in their respective $\pm 1\sigma$ intervals, while 15 of the 20 parameters in the model three oval hot spots fall in their respective $\pm 1\sigma$ intervals. All of the parameters in the model with two oval hot spots fall in their respective $\pm 2\sigma$ intervals, and 19 of the 20 parameters in the model with three oval hot spots fall in their respective $\pm 2\sigma$ intervals. Thus, the best-fit models of \citet{miller19} easily pass this posterior-predictive assessment, which assumes that the realizations of the synthetic data generated using the best-fit two- and three-spot models of \citet{miller19} exactly match the actual PSR~J0030$+$0451 data collected by \textit{NICER} and then analyzes that ``synthetic'' data using the models developed by \citet{miller19}.

\section{Dependence of $M$ and $R$ Constraints on Spin Rate and Hot Spot Size}
\label{sec:model_properties}

As discussed in several previous works \citep[see, e.g.,][]{miller98,poutanen06,psaltis14a}, the precision with which $M$ and $R_{\rm eq}$ can be determined depends on many properties of the pulsar system. In particular, the colatitudes of hot spots and the angle between the sightline to the observer and the stellar spin axis strongly affect the precision. Geometries in which the spots and the sightline to the observer are near the plane of the star's rotational equator provide tighter constraints than geometries in which either or both are close to the star's spin axis \citep[see, e.g.,][]{bogdanov08,lo13,miller15}. In this section, we illustrate how the constraints on $M$ and $R_{\rm eq}$ depend on the spin rate of the pulsar and the size of the hot spot by analyzing a set of synthetic pulse profiles produced by pulsar models with a single, circular, uniform-temperature hot spot and a phase-independent background, but different spin rates and spot diameters.

\begin{figure}
\centering
\includegraphics[clip,trim = 0cm 0cm 0cm 0cm,width=0.7\textwidth]{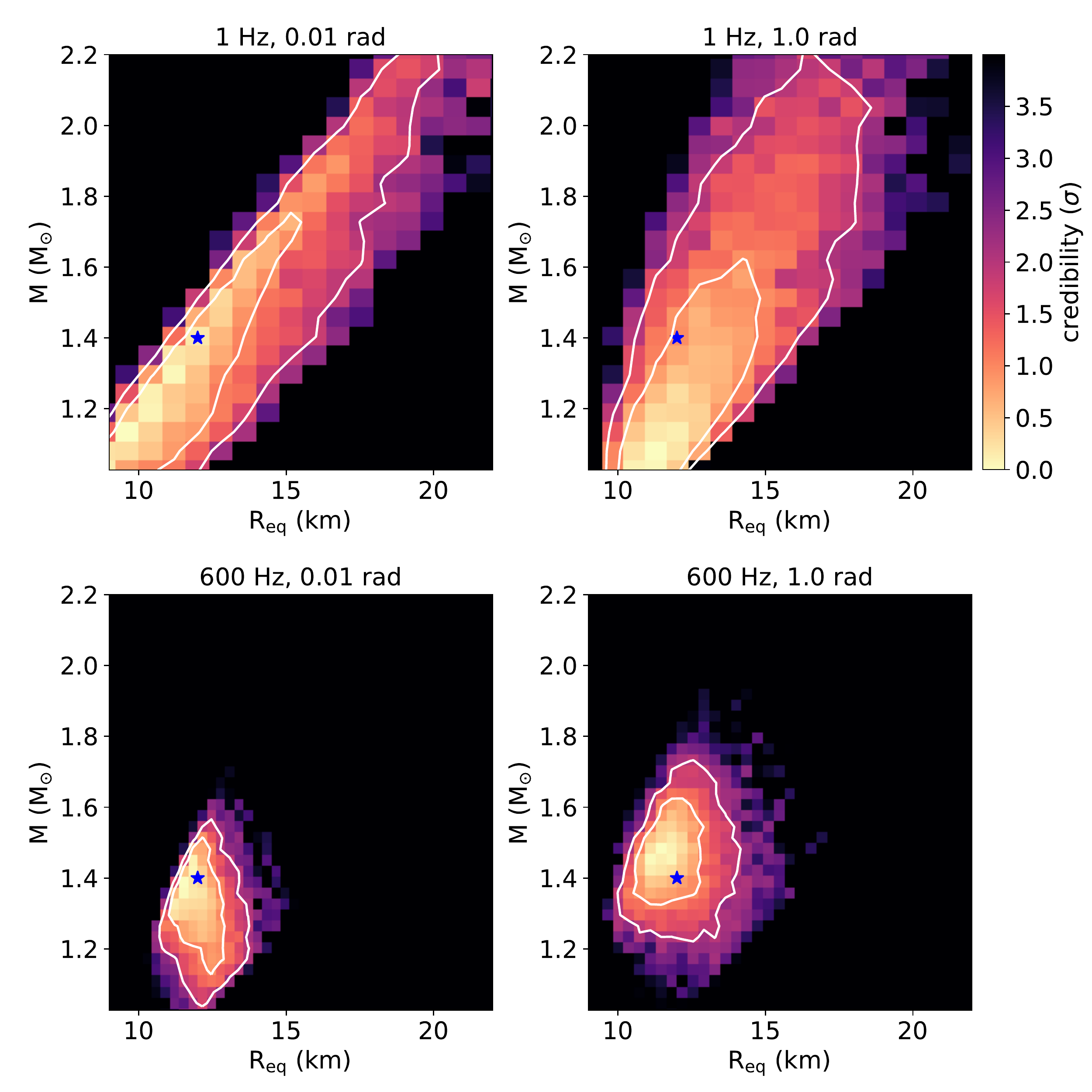}
\caption{
Posterior probability distributions in the $M$--$R_{\rm eq}$ plane obtained by analyzing synthetic \textit{NICER} data from a pulsar with a single hot spot (see text), using the Illinois-Maryland code. The distributions are colored by credibility interval, as explained in Sections~\ref{sec:onespot} and~\ref{sec:twospot}. The blue star marks the $M$ and $R_{\rm eq}$ values assumed in generating the synthetic data; the white contours enclose the 1$\sigma$ and 2$\sigma$ credible regions. The four sets of synthetic pulse profile data that were analyzed were generated using the spin-frequency and spot-size combinations listed at the top of each panel. The panels are referred to in the text as Case~1 (top left), Case~2 (top right), Case~3 (bottom left), and Case~4 (bottom right). Note that the stellar compactness ($M/R_{\rm eq}$) is not always the best-determined stellar property; indeed, for Cases~3 and~4, the radius is determined with a better fractional precision than the compactness.
\label{fig:four_synth}}
\end{figure}

To illustrate the strong dependence of the precisions of $M$ and $R_{\rm eq}$ estimates on the spin rate and their much weaker dependence on the size of the hot spot, we generated synthetic \textit{NICER} pulse profile data using pulsar models with the following four combinations of spin frequency $\nu_{\rm rot}$ and hot spot angular radius $\Delta\theta$, namely,
Case~1: $\nu_{\rm rot}=1$~Hz, $\Delta\theta = 0.01$~rad; 
Case~2: $\nu_{\rm rot}=1$~Hz, $\Delta\theta = 1.0$~rad; 
Case~3: $\nu_{\rm rot}=600$~Hz, $\Delta\theta=0.01$~rad; and 
Case~4: $\nu_{\rm rot}=600$~Hz, $\Delta\theta = 1.0$~rad.
The values of the other parameters in the pulsar model were kept the same for all four cases. These parameters and their values were: stellar equatorial radius $R_{\rm eq}=12$~km, stellar mass $M=1.4\,$M$_{\odot}$, angle between the sightline to the observer and the spin axis $\theta_{\rm obs}=1.047198$\,rad ($60^{\circ}$), colatitude of the center of the hot spot $\theta_c=1.745329$~rad ($100^{\circ}$), effective temperature of the hot spot $kT_{\rm eff}=0.231139$~keV, distance to the pulsar $D=0.2$~kpc, and hydrogen column density $N_{\rm H}=1.0\times 10^{20}$~cm$^{-2}$. For all four synthetic pulse profiles, we set the expected total number of counts from the hot spot to $\sim\,$360,000 and the expected total number of background counts to $\sim\,$1,000,000, approximately the number of counts from the hot spots and from the phase-independent background obtained by \citet{miller19} in the best fit of their model with three oval spots to the \textit{NICER} pulse profile data on PSR~J0030$+$0451. We used a spectrum for the synthetic background that matches the spectrum of the phase-independent background found by \citet{miller19} in their best fit to the PSR~J0030$+$0451 data.

Figure~\ref{fig:four_synth} shows the posterior distributions in the $M$--$R_{\rm eq}$ plane obtained by analyzing these four sets of synthetic data using the parameter-estimation algorithm developed by \citet{miller19} and a pulse profile model with a single, circular, uniform temperature hot spot. These plots demonstrate how precisely the mass and radius of the star are encoded in the observed pulse profile by special and general relativistic effects. 

If the star is rotating very slowly, Doppler boosts, aberration, and other special relativistic effects are unimportant, but gravitational lensing by the star---which depends only on $M/R_{\rm eq}$---affects the pulse profile. Hence, fitting pulse profile models to the observed pulse profile of a slowly rotating star is expected to place a tighter constraint on $M/R_{\rm eq}$ than on $M$ and $R_{\rm eq}$ separately. This is illustrated by Case~1, which assumes a very slow (1~Hz) stellar rotation rate and a very small spot ($\Delta\theta = 0.01$~rad); in this case, the peak of the probability density roughly follows the expected $M/R_{\rm eq} = 0.11\,M_\odot\,{\rm km}^{-1}$ relation at low masses and radii, but with a significant uncertainty in the slope at low masses and radii and a much larger uncertainty in the slope at high masses and radii. Case~2 also assumes a 1~Hz rotation rate but a much larger spot ($\Delta\theta = 1.0$~rad); in this case, the $1\sigma$ credible region is substantially larger than for a very small spot and the probability density peak does not follow a line with $M/R_{\rm eq} \approx {\rm const.}$  A spot this large is only gradually occulted by the body of the star as the star rotates, which probably explains why the slope of the probability density peak is steeper. In both cases, fitting the observed pulse profile constrains $R_{\rm eq}$, but not very tightly.

If instead the star is rotating fairly rapidly ($\nu_{\rm rot} \gtrsim 300$~Hz), the special relativistic effects produced by the velocity of the star's emitting surface relative to the observer---which is proportional to $R_{\rm eq}$---affects the pulse profile. Hence, fitting pulse profile models to the observed pulse profile of a rapidly rotating star not only constrains $M/R_{\rm eq}$ but also constrains $M$ and $R_{\rm eq}$ separately. This can be seen by comparing Cases~1 and~2, which assume a very slow (1~Hz) rotation rate, with Cases~3 and~4, which assume the same spot sizes as Cases~1 and~2 but a much faster (600~Hz) rotation rate.\footnote{Using the OS approximation typically introduces errors of several percent for stellar spin frequencies $\sim\,$600~Hz, as explained in Paper~II. However, errors of this size are unimportant for this purely illustrative comparison.}  As Case~3 shows, for some rotation rates and spot sizes, the fractional uncertainty in the estimate of $R_{\rm eq}$ can be smaller than the fractional uncertainty in the estimate of $M/R_{\rm eq}$ (see also \citealt{psaltis14a,miller15,2016ApJ...833..244S}).

As noted above, the shape of the observed pulse profile also depends on the size of the hot spot. Radiation from a larger hot spot will be observed over a wider range of rotational phases, blurring the effects produced by the Doppler boost, aberration, and self-occultation of the hot spot by the star. We therefore expect that for a given number of counts from the spot and a given number of background counts, the credible region in the $M$--$R_{\rm eq}$ plane defined by a given cumulative probability will be somewhat larger, in general, for a larger hot spot. This is illustrated by Case~1 compared to Case~2 and Case~3 compared to Case~4. For all four synthetic data sets analyzed to produce Figure~\ref{fig:four_synth}, the number of counts from the spot and from the background were assumed to be the same, even though the larger spots have areas $10^4$ times greater than the smaller spots. If all these pulsars were observed for approximately the same length of time and all the spots had the same temperature, $\sim 10^4$ times as many counts would be collected from the pulsars with the larger spots than from the pulsars with the smaller spots. If all else was equal, this would cause the constraints on $M$ and $R_{\rm eq}$ to be $\sim 10^2$ times tighter for the stars with the larger spots than for the stars with the smaller spots.

\section{Future Exploration of Possible Systematic Effects}\label{sec:systematics}

Our simple description of hot spots cannot be entirely accurate.  For example, the hotter regions on neutron stars will not be circles or ovals of perfectly uniform temperature.  However, for the purpose of measuring the masses and radii of neutron stars observed using \textit{NICER} the models can be adequate without being accurate in every detail.  

Indeed, \citet{lo13} and \citet{miller15} have investigated the effects of systematic modeling errors on estimates of $M$ and $R_{\rm eq}$ obtained by fitting X-ray pulse profile models to observations of the thermonuclear X-ray burst oscillations produced by some accreting MSPs or the X-ray pulsations produced by hot spots on the surfaces of non-accreting MSPs. These investigations used pulse profile models to analyze synthetic pulse profiles in which the hot spot contributes a total of $10^6$ counts and an unmodulated background contributes an additional $9\times10^6$ counts. This number of hot-spot counts is comparable to that which could be obtained by observing the brightest non-accreting MSPs using \textit{NICER} for about $10^6$~seconds on-source (corresponding roughly to exposure accumulated over a year of calendar time; see \citealt{bogdanov19a}), or by combining observations of the oscillations of several dozen relatively bright X-ray bursts from the same accreting MSP (see \citealt{lo13}). This assumed ratio of oscillating counts to background counts is conservative compared with the \textit{NICER} data on PSR~J0030$+$0415, for which the ratio of oscillating counts to background counts through January 2019 (1.8 million seconds of observing time) was about 360,000 to 1,000,000, rather than 1:9.

Importantly, these studies demonstrated that fitting rotating hot spot models to energy-dependent X-ray pulse profiles gives results that are robust against several types of systematic effects. In particular, when pulse profile models were fit to synthetic data generated using spot shapes, energy spectra, temperature gradients, or surface beaming patterns different from those assumed in the models, the fits did not simultaneously produce (i)~a statistically good fit (as indicated by the value of $\chi^2$ obtained by comparing the best-fit bolometric pulse profile model with the bolometric pulse profile data), (ii)~apparently tight constraints on the stellar mass and radius, and (iii)~a significant bias in their inferred values. Thus, at least for these modeling errors, pulse profile analyses did not yield tight but misleading constraints.  

There are other possible discrepancies between the model pulse profile used to analyze the observed X-ray pulse profile of a rotation-powered MSP and the actual profile of the pulsar.  It would be worthwhile to investigate further their effects on mass and radius estimates to determine whether or not they introduce significant biases.  Possible discrepancies include:

\textit{Chemical composition of the atmosphere.}---The cosmic abundance of hydrogen, and the expected rapid sinking of heavier elements in neutron star atmospheres, justifies our standard assumption that the topmost layer of the neutron star atmosphere is pure hydrogen. In \citet{miller19}, the \textit{NICER} data from PSR~J0030+0451 were also fit using a model with a non-magnetic NSX helium atmosphere \citep{holai01}. This analysis resulted in a best-fit mass of $2.7~M_\odot$, with most of the posterior probability above $2.0~M_\odot$. The tendency towards implausibly large NS masses suggests that a He atmosphere is unlikely, but this deserves further exploration, not only for PSR~J0030+0451 but also for other MSPs that are \textit{NICER} targets.

We also note that, since we are able to distinguish between helium and hydrogen atmosphere model results, which are caused by beaming patterns that differ at $\le5$\% in the appropriate energy range (see Section~\ref{sec:atmosphere}), we expect to be sensitive to other atmosphere model effects which cause greater than  $1-2$\% differences in beaming patterns.

\textit{Magnetic fields.}---As discussed in Section~\ref{sec:atmosphere}, for the surface magnetic field strength inferred assuming a centered, purely dipolar stellar field for PSR J0030+0415, the effects of the magnetic field on the spectrum and beaming pattern of the radiation from the hot spots are negligible.

\citet{2021ApJ...907...63K} investigated offset dipole plus quadrupole magnetic field configurations using static vacuum field and force-free global magnetosphere models aimed at reproducing the hot spot patterns inferred for PSR J0030+0451 by \citet{miller19} and \citet{riley19}. They demonstrated that by assuming the hot spots correspond to the open field line regions on the surface (i.e., the pulsar magnetic polar caps) the seemingly unusual spot geometries inferred for this pulsar from the \textit{NICER} data can be reproduced very well by a number of plausible dipole plus quadrupole magnetic field arrangements. In addition, despite the presence of multipoles, the field is found to rarely exceed $\sim$10 times the dipole spin field, and is thus typically $\lesssim B_0$. Therefore, in the case of PSR J0030+0451 magnetic influences on radiation transport in the neutron star atmosphere are likely small in the \textit{NICER} band. While strong magnetic fields are likely to produce beaming patterns that are incompatible with the energy-dependent pulse profile from PSR J0030+0451, the effect of different atmosphere beaming patterns on the NS mass and radius inferences deserves further investigation.

\textit{Depth of energy deposition.}---The surface thermal emission from rotation-powered MSPs is most likely caused by heating by backward flowing energetic particles produced by pair cascades at the polar caps or in the outer magnetosphere, and by global return currents\footnote{These pulsars have characteristic ages much greater than the age $\sim 10^6$\,yr when the emission from the surface of a neutron star produced by its initially hot interior could exceed the emission produced by backward flowing particles and return currents.} \citep[e.g.,][]{2006ApJ...648L..51S,2011ApJ...743..181H}. In our models, we assume that the energy of the returning particles that bombard the neutron star surface is deposited many optical depths below the photosphere. To assess the validity of this assumption, note that plasma instabilities are unlikely to play a significant role in the stopping process (see, e.g., \citealt{2013MNRAS.429...20T}), that a returning electron or positron with a Lorentz factor $\gamma>10^3$ has a total stopping depth\footnote{\url{https://physics.nist.gov/PhysRefData/Star/Text/ESTAR.html}} in hydrogen larger than 90~g~cm$^{-2}$, and that it will therefore deposit the bulk of its energy at a density around $\rho\sim 10^3$~g~cm$^{-3}$ \citep{Bethe34,Tsai74}. Given that the column depth of a typical nonmagnetic NS H atmosphere is 0.1--1~g~cm$^{-2}$ (see, e.g., \citealt{zavlinetal96}), a particle with a Lorentz factor $\gamma>10^3$ will deposit negligible energy in such an atmosphere. Lower-energy particles, with $\gamma=10-10^3$, would stop at depths of 2--90~g~cm$^{-2}$ in H and therefore could potentially deposit some of their energy at densities as low as 10~g~cm$^{-3}$. In strong magnetic fields, the stopping depth of electrons is reduced by magneto-Coulomb interactions in which the electrons are excited to higher Landau levels in the Coulomb fields of ions and then de-excite by emitting synchrotron or cyclotron radiation \citep{Ho07,potekhin14}. However, the MSPs that are \textit{NICER} targets are thought to have surface magnetic fields too weak for this process to significantly decrease the stopping depths of electrons or positrons.

\citet{baubock19} investigated the radiative properties of a neutron star atmosphere (assumed to be gray) heated by returning particles that deposit their energy in the upper layers of the atmosphere. They find that this occurs for low-energy particles with $\gamma\sim2$--10.  \citet{Salmi2020} performed a similar analysis but treated Compton scattering within the atmosphere exactly. They conclude that if most of the returning particles have Lorentz factors $\gamma\lesssim100$, their energy will be deposited at relatively low optical depths, causing the atmosphere to develop a hotter, optically thin skin on top of a cooler, optically thick region. 

\citet{2013MNRAS.429...20T} find that most of the heating of the atmosphere by returning particles is caused by those with $\gamma\gg 10$--100, and in some cases as high as $\sim 10^7$ (see especially Figures 17 and 18 in \citealt{2013MNRAS.429...20T}).  For MSPs in particular, the lowest energy particles in calculated pair spectra (see \citealt{2011ApJ...743..181H}, Fig.~10) have $\gamma \sim 10^3 - 10^4$.  This is because the very low magnetic fields of MSPs require much higher photon energies to create pairs.   Even though for the time-dependent cascades there are some low energy thermal particles moving downward, their number densities are 100 to 1000 times lower than the higher energy pairs, so they are unlikely to contribute much of the heating.  It therefore seems probable that our standard assumption that the energy of returning particles is deposited at large optical depths is correct for the MSPs observed with \textit{NICER}. Nevertheless, it is important to examine the effect that small deviations in the atmosphere model beaming pattern might have on the inferred values of the NS parameters.

\textit{Non-spot background emission}---For the \textit{NICER} MSP data under consideration for $M-R$ parameter estimation analyses, a substantial fraction of the observed emission does not originate from the hot regions on the stellar surface and is not modulated at the pulsar spin period. As described in Paper I, there are a number of sources of background radiation that can affect \textit{NICER} X-ray data, which include particle and photon background from the local low-Earth-orbit environment of \textit{NICER}, the unresolved diffuse X-ray background, other point sources that fall within the telescope field of view, and other emission processes in the vicinity of the pulsar under consideration. In the analyses of the \textit{NICER} data of PSR~J0030$+$0451 presented in \cite{miller19} and \cite{riley19}, no estimated background is explicitly taken into account; this is the most conservative approach. Instead, non-hot-spot emission in each detector channel is treated as a free parameter and is assumed to have no dependence on spin phase. Thus, all background emission that is not modulated at the pulsar period is incorporated in a straightforward fashion. 

It is important the emphasize that although such an assumption introduces a large number of additional free parameters, this would not result in bias as the best solution is still reachable by the model. Instead, it would lead to greater imprecision in the measurement, i.e., broader credible intervals for the inferred parameters compared to a model that can accurately describe the observed pulse profile but contains a smaller number of parameters. This is demonstrated by the analysis in \citet{miller19} of the synthetic \textit{NICER} data that mimicked that of PSR J0030+0451 (including the estimated background level), which also assumed no knowledge of the background emission and produced estimated values and credible regions consistent with the values assumed in constructing the synthetic pulse profile.

By the same token, the inclusion of independent estimates of the non-spot contribution to the observed X-ray emission as a function of energy in the parameter estimation analyses can potentially lead to significantly tighter constraints on the NS mass and radius estimates, provided that the background estimates are reliable and do not introduce measurement bias by over/underestimating the true background. As our understanding of instrumental and other sources of background improves, we anticipate exploiting such knowledge as informative priors in our Bayesian inferences, with potentially significant implications for the precision and accuracy of the resulting NS mass and radius estimates.
As noted in Section\ref{sec:backgrounds}, we are developing the capability to incorporate omplementary spectral information about the non-source emission that accounts for a substantial fraction of the events collected in \textit{NICER} observations of the MSP targets under consideration. This includes using the background estimated from the ``3c50" and space weather-based techniques developed by the \textit{NICER} team \citep[see][]{bogdanov19a}, as well as from existing X-ray imaging data from other observatories. The details of the implementation of background information will be described in subsequent publications.

\textit{Pulsed non-thermal emission.}--- While all backgrounds that are not modulated at the rotation frequency of the pulsar are included in the analysis algorithms being used by \citet{miller19} and \citet{riley19}, any modulated non-thermal emission from the pulsar in the energy band that is currently being analyzed (0.3--3 keV) is not included in our models. For PSR J0437--4715, there is marginal evidence for pulsations in the 2--20 keV band \citep{guillot16}, but the photon flux from this apparently non-thermal emission is only $\sim 3$\% of the total flux at the energies $<2$~keV where the modulated thermal component is prominent. PSR J0030$+$0451 also exhibits a power-law spectral tail evident above $\sim$2 keV in \textit{XMM-Newton} data \citep{bogdanov09} but is too faint to test for the presence of pulsations.
It is therefore unlikely that this apparently non-thermal pulsed flux significantly affected our analysis of the pulsed thermal emission from PSR J0030$+$0451, but additional studies aimed at quantifying its impact on the NS mass and radius estimates are needed.

\textit{NICER instrumental calibration.}---Because the energy-dependent sensitivities of the \textit{NICER} detectors are not perfectly known\footnote{Due to the challenge of obtaining suitable reference targets for precise calibration measurements, all observations with X-ray telescopes suffer from an uncertainty in the absolute effective area across all energies, which is commonly estimated to be at the level of $\pm$10\%.}, uncertainties in their responses could in principle affect estimates of the masses and radii of MSPs made using \textit{NICER} data.  Energy-independent uncertainties in the effective area were already incorporated in \citet{miller19} by adding extra effective uncertainty to the distance to the MSPs being observed, and in \citet{riley19} by inclusion of calibration parameters (estimated via a procedure similar to that described by \citealt{ludlam2018} using observations of the Crab Nebula as a reference).  The effect of energy-dependent uncertainties in the response of the \textit{NICER} XTI could be more subtle.   However, using different methods (see the individual papers for details), \citet{miller19} and \citet{riley19} found no evidence that systematic errors in the instrumental calibration biased their estimates of the mass and radius of PSR~J0030+0451. The \textit{NICER} calibration is being refined with time, so the evolution of mass-radius estimates as updated calibrations are used will provide information about whether calibration errors could significantly affect these estimates.

\section{Conclusions} 
\label{sec:discussion}
We have described the physical components of the model of pulsed thermal X-ray emission by MSPs that we have used and will continue to use to analyze \textit{NICER} data on rotation-powered MSPs to infer their masses and radii. We also reported a comparison of model pulse profiles produced by a star with $R_{\rm eq}/M \approx 3$, performed to cross-check the accuracy of these codes for future analyses of \textit{NICER} pulse profile data, where considering more compact stars might be relevant.

We have also described our verification of the parameter estimation algorithms and codes used by \citet{miller19} and \citet{riley19} to analyze the \textit{NICER} data on PSR~J0030$+$0451 by using them to analyze synthetic pulse profile data produced by two pulsar models, one with a single hot spot and one with two hot spots. We showed that these codes yield credible regions for all the parameters in the pulse-profile models used to analyze the synthetic data that are consistent with each other and with the parameter values used to generate the synthetic pulse profile data. The \textit{NICER} synthetic data sets and auxiliary files used in these comparisons are provided to the community on Zenodo as supplementary material to this article to allow testing of other independently developed codes. 

We also provided an illustration of how the stellar spin rate and the size of its hot spot affect the constraints on the mass and radius of a neutron stars that can be obtained using the pulse-profile fitting method. The results obtained by using this technique to analyze synthetic \textit{NICER} pulse profiles generated using four model pulsars highlight the importance of rapid spin for obtaining constraints on $M$ and $R_{\rm eq}$ separately. In the case of hot spot size, for equivalent source and background counts, neutron stars with smaller spots would, in principle, produce tighter $M$ and $R_{\rm eq}$ constraints, but in practice would require substantially longer exposures compared to stars with larger spots to achieve this.

We discussed the assumptions made in our analyses concerning the composition of the neutron star atmosphere, the strength and geometry of its surface magnetic field, and the depth at which returning particles deposit their energy.  We concluded that these assumptions are reasonable and based on our current best understanding of the surface properties of rotation-powered MSPs as informed by multi-wavelength observations and theory. Nevertheless, further study of the aforementioned effects may be useful to ascertain their impact on the neutron star mass and radius estimates. Finally, we argued that the estimates of the mass and radius of PSR~J0030$+$0451 that have been made using \textit{NICER} are not significantly affected by the current approach to modeling the unpulsed background, the possible pulsation of its non-thermal medium-energy X-ray emission, and that instrumental calibration errors are unlikely to introduce significant biases in our mass and radius estimates.  We concluded that these assumptions are likely accurate, though more exhaustive studies of some may be useful.

The results presented here provide additional confidence that the current measurements of the mass and radius of PSR~J0030$+$0451 and future measurements of the masses and radii of this MSP and others made using \textit{NICER} will be reliable as well as precise.

\acknowledgments
This work was supported by NASA through the \textit{NICER} mission and the Astrophysics Explorers Program. S.B.~was funded in part by NASA grants NNX17AC28G and 80NSSC20K0275. W.C.G.H.~appreciates use of computer facilities at the Kavli Institute for Particle Astrophysics and Cosmology and acknowledges support through grant 80NSSC20K0278 from NASA. A.J.D., M.C.M., and F.K.L.~acknowledge the University of Maryland supercomputing resources that were made available for conducting the research reported in this Letter.  M.C.M.~thanks the Radboud Excellence Initiative for supporting his stay at Radboud University, and was also supported by a Visiting Researcher position at Perimeter Institute for Theoretical Physics in the early stages of this project. T.E.R., D.C., and A.L.W. acknowledge support from ERC Starting Grant No.~639217 CSINEUTRONSTAR and ERC Consolidator Grant No.~865768 AEONS. This work was sponsored by NWO Exact and Natural Sciences for the use of supercomputer facilities, and was carried out on the Dutch national e-infrastructure with the support of SURF Cooperative. Z.W. acknowledges support from the NASA postdoctoral program. S.G. acknowledges the support of the CNES. S.M.M.~acknowledges support from NSERC. T.E.R, D.C., and A.L.W. are grateful to W.~M.~Farr and I.~Mandel for introducing us to the notion of simulation-based calibration of posterior computation, and for their advice and suggestions about weaker cross-checks, all of which informed our perspective on what level of calibration we attained. This research has made use of data products and software provided by the High Energy Astrophysics Science Archive Research Center (HEASARC), which is a service of the Astrophysics Science Division at NASA/GSFC and the High Energy Astrophysics Division of the Smithsonian  Astrophysical Observatory.  We acknowledge extensive use of NASA's Astrophysics Data System (ADS) Bibliographic Services and the ArXiv. 

\facilities{\textit{NICER}}

\software{Python/C language \citep{4160250}, GNU Scientific Library \citep{gough2009gnu}, Cython  \citep{5582062}, SciPy \citep{2020SciPy-NMeth}, NumPy \citep{2011CSE....13b..22V}, Matplotlib \citep{4160265}, MultiNest \citep{2009MNRAS.398.1601F}, emcee \citep{2013PASP..125..306F}, GetDist \citep{Lewis2019}}

\bibliographystyle{aasjournal}
\bibliography{wp_references}

\begin{thebibliography}{}
\expandafter\ifx\csname natexlab\endcsname\relax\def\natexlab#1{#1}\fi
\providecommand{\url}[1]{\href{#1}{#1}}

\bibitem[{{Alcock} \& {Illarionov}(1980)}]{alcockillarionov80}
{Alcock}, C., \& {Illarionov}, A. 1980, \apj, 235, 534

\bibitem[{{AlGendy} \& {Morsink}(2014)}]{2014ApJ...791...78A}
{AlGendy}, M., \& {Morsink}, S.~M. 2014, \apj, 791, 78

\bibitem[{{Arnaud}(1996)}]{arnaud96}
{Arnaud}, K.~A. 1996, in Astronomical Society of the Pacific Conference Series,
  Vol. 101, Astronomical Data Analysis Software and Systems V, ed. G.~H.
  {Jacoby} \& J.~{Barnes}, 17--+

\bibitem[{{Badnell} {et~al.}(2005){Badnell}, {Bautista}, {Butler}, {Delahaye},
  {Mendoza}, {Palmeri}, {Zeippen}, \& {Seaton}}]{badnelletal05}
{Badnell}, N.~R., {Bautista}, M.~A., {Butler}, K., {et~al.} 2005, \mnras, 360,
  458

\bibitem[{{Baub{\"o}ck} {et~al.}(2019){Baub{\"o}ck}, {Psaltis}, \&
  {{\"O}zel}}]{baubock19}
{Baub{\"o}ck}, M., {Psaltis}, D., \& {{\"O}zel}, F. 2019, \apj, 872, 162

\bibitem[{{Baym} {et~al.}(2018){Baym}, {Hatsuda}, {Kojo}, {Powell}, {Song}, \&
  {Takatsuka}}]{Baym18}
{Baym}, G., {Hatsuda}, T., {Kojo}, T., {et~al.} 2018, Reports on Progress in
  Physics, 81, 056902

\bibitem[{{Behnel} {et~al.}(2011){Behnel}, {Bradshaw}, {Citro}, {Dalcin},
  {Seljebotn}, \& {Smith}}]{5582062}
{Behnel}, S., {Bradshaw}, R., {Citro}, C., {et~al.} 2011, Computing in Science
  Engineering, 13, 31

\bibitem[{{Beloborodov}(2002)}]{beloborodov02}
{Beloborodov}, A.~M. 2002, \apjl, 566, L85

\bibitem[{{Berry} {et~al.}(2015){Berry}, {Mandel}, {Middleton}, {Singer},
  {Urban}, {Vecchio}, {Vitale}, {Cannon}, {Farr}, {Farr}, {Graff}, {Hanna},
  {Haster}, {Mohapatra}, {Pankow}, {Price}, {Sidery}, \& {Veitch}}]{Berry2015}
{Berry}, C. P.~L., {Mandel}, I., {Middleton}, H., {et~al.} 2015, \apj, 804, 114

\bibitem[{{Betancourt}(2017)}]{Betancourt2017}
{Betancourt}, M. 2017, arXiv e-prints, arXiv:1701.02434

\bibitem[{{Bethe} \& {Heitler}(1934)}]{Bethe34}
{Bethe}, H., \& {Heitler}, W. 1934, Proceedings of the Royal Society of London
  Series A, 146, 83

\bibitem[{{Bildsten} {et~al.}(1992){Bildsten}, {Salpeter}, \&
  {Wasserman}}]{1992ApJ...384..143B}
{Bildsten}, L., {Salpeter}, E.~E., \& {Wasserman}, I. 1992, \apj, 384, 143

\bibitem[{{Bilous} {et~al.}(2019){Bilous}, {Watts}, {Harding}, \&
  E.}]{bilous19}
{Bilous}, A.~V., {Watts}, A.~L., {Harding}, A.~K., \& E., R.~T. 2019, \apjl,
  887, L23

\bibitem[{{Blaes} {et~al.}(1992){Blaes}, {Blandford}, {Madau}, \&
  {Yan}}]{blaesetal92}
{Blaes}, O.~M., {Blandford}, R.~D., {Madau}, P., \& {Yan}, L. 1992, \apj, 399,
  634

\bibitem[{{Bogdanov} \& {Grindlay}(2009)}]{bogdanov09}
{Bogdanov}, S., \& {Grindlay}, J.~E. 2009, \apj, 703, 1557

\bibitem[{{Bogdanov} {et~al.}(2008){Bogdanov}, {Grindlay}, \&
  {Rybicki}}]{bogdanov08}
{Bogdanov}, S., {Grindlay}, J.~E., \& {Rybicki}, G.~B. 2008, \apj, 689, 407

\bibitem[{{Bogdanov} {et~al.}(2019{\natexlab{a}}){Bogdanov}, {Guillot}, {Ray},
  T., D., G., \& M.}]{bogdanov19a}
{Bogdanov}, S., {Guillot}, S., {Ray}, P., {et~al.} 2019{\natexlab{a}}, \apjl,
  887, L25

\bibitem[{{Bogdanov} {et~al.}(2016){Bogdanov}, {Heinke}, {{\"O}zel}, \&
  {G{\"u}ver}}]{2016ApJ...831..184B}
{Bogdanov}, S., {Heinke}, C.~O., {{\"O}zel}, F., \& {G{\"u}ver}, T. 2016, \apj,
  831, 184

\bibitem[{{Bogdanov} {et~al.}(2019{\natexlab{b}}){Bogdanov}, {Lamb},
  {Mahmoodifar}, C., M., \& E.}]{bogdanov19b}
{Bogdanov}, S., {Lamb}, F.~K., {Mahmoodifar}, S., {et~al.} 2019{\natexlab{b}},
  \apjl, 887, L26

\bibitem[{{Braje} {et~al.}(2000){Braje}, {Romani}, \& {Rauch}}]{braje00}
{Braje}, T.~M., {Romani}, R.~W., \& {Rauch}, K.~P. 2000, \apj, 531, 447

\bibitem[{{Brewer} \& {Foreman-Mackey}(2016)}]{2016arXiv160603757B}
{Brewer}, B.~J., \& {Foreman-Mackey}, D. 2016, ArXiv e-prints, arXiv:1606.03757

\bibitem[{{Brown} {et~al.}(2002){Brown}, {Bildsten}, \& {Chang}}]{Brown02}
{Brown}, E.~F., {Bildsten}, L., \& {Chang}, P. 2002, \apj, 574, 920

\bibitem[{{Buchdahl}(1959)}]{1959PhRv..116.1027B}
{Buchdahl}, H.~A. 1959, Physical Review, 116, 1027

\bibitem[{{Cadeau} {et~al.}(2007){Cadeau}, {Morsink}, {Leahy}, \&
  {Campbell}}]{cadeau07}
{Cadeau}, C., {Morsink}, S.~M., {Leahy}, D., \& {Campbell}, S.~S. 2007, \apj,
  654, 458

\bibitem[{{Chang} \& {Bildsten}(2003)}]{changbildsten03}
{Chang}, P., \& {Bildsten}, L. 2003, \apj, 585, 464

\bibitem[{{Chang} \& {Bildsten}(2004)}]{2004ApJ...605..830C}
---. 2004, \apj, 605, 830

\bibitem[{Chen {et~al.}(2020)Chen, Yuan, \& Vasilopoulos}]{chen2020}
Chen, A.~Y., Yuan, Y., \& Vasilopoulos, G. 2020, The Astrophysical Journal,
  893, L38.
\newblock \url{https://doi.org/10.3847%2F2041-8213%2Fab85c5}

\bibitem[{{Colgan} {et~al.}(2016){Colgan}, {Kilcrease}, {Magee}, {Sherrill},
  {Abdallah}, {Hakel}, {Fontes}, {Guzik}, \& {Mussack}}]{colganetal16}
{Colgan}, J., {Kilcrease}, D.~P., {Magee}, N.~H., {et~al.} 2016, \apj, 817, 116

\bibitem[{{Contopoulos} {et~al.}(1999){Contopoulos}, {Kazanas}, \&
  {Fendt}}]{1999ApJ...511..351C}
{Contopoulos}, I., {Kazanas}, D., \& {Fendt}, C. 1999, \apj, 511, 351

\bibitem[{{Contopoulos} \& {Spitkovsky}(2006)}]{2006ApJ...643.1139C}
{Contopoulos}, I., \& {Spitkovsky}, A. 2006, \apj, 643, 1139

\bibitem[{{Feroz} {et~al.}(2009){Feroz}, {Hobson}, \&
  {Bridges}}]{2009MNRAS.398.1601F}
{Feroz}, F., {Hobson}, M.~P., \& {Bridges}, M. 2009, \mnras, 398, 1601

\bibitem[{{Foreman-Mackey} {et~al.}(2013){Foreman-Mackey}, {Hogg}, {Lang}, \&
  {Goodman}}]{2013PASP..125..306F}
{Foreman-Mackey}, D., {Hogg}, D.~W., {Lang}, D., \& {Goodman}, J. 2013, \pasp,
  125, 306

\bibitem[{{G{\"a}nsicke} {et~al.}(2002){G{\"a}nsicke}, {Braje}, \&
  {Romani}}]{gansickeetal02}
{G{\"a}nsicke}, B.~T., {Braje}, T.~M., \& {Romani}, R.~W. 2002, \aap, 386, 1001

\bibitem[{{Gendreau} {et~al.}(2016){Gendreau}, {Arzoumanian}, {Adkins},
  {Albert}, {Anders}, {Aylward}, {Baker}, {Balsamo}, {Bamford}, {Benegalrao},
  {Berry}, {Bhalwani}, {Black}, {Blaurock}, {Bronke}, {Brown}, {Budinoff},
  {Cantwell}, {Cazeau}, {Chen}, {Clement}, {Colangelo}, {Coleman},
  {Coopersmith}, {Dehaven}, {Doty}, {Egan}, {Enoto}, {Fan}, {Ferro}, {Foster},
  {Galassi}, {Gallo}, {Green}, {Grosh}, {Ha}, {Hasouneh}, {Heefner}, {Hestnes},
  {Hoge}, {Jacobs}, {J{\o}rgensen}, {Kaiser}, {Kellogg}, {Kenyon}, {Koenecke},
  {Kozon}, {LaMarr}, {Lambertson}, {Larson}, {Lentine}, {Lewis}, {Lilly},
  {Liu}, {Malonis}, {Manthripragada}, {Markwardt}, {Matonak}, {Mcginnis},
  {Miller}, {Mitchell}, {Mitchell}, {Mohammed}, {Monroe}, {Montt de Garcia},
  {Mul{\'e}}, {Nagao}, {Ngo}, {Norris}, {Norwood}, {Novotka}, {Okajima},
  {Olsen}, {Onyeachu}, {Orosco}, {Peterson}, {Pevear}, {Pham}, {Pollard},
  {Pope}, {Powers}, {Powers}, {Price}, {Prigozhin}, {Ramirez}, {Reid},
  {Remillard}, {Rogstad}, {Rosecrans}, {Rowe}, {Sager}, {Sanders}, {Savadkin},
  {Saylor}, {Schaeffer}, {Schweiss}, {Semper}, {Serlemitsos}, {Shackelford},
  {Soong}, {Struebel}, {Vezie}, {Villasenor}, {Winternitz}, {Wofford},
  {Wright}, {Yang}, \& {Yu}}]{2016SPIE.9905E..1HG}
{Gendreau}, K.~C., {Arzoumanian}, Z., {Adkins}, P.~W., {et~al.} 2016, in
  \procspie, Vol. 9905, Space Telescopes and Instrumentation 2016: Ultraviolet
  to Gamma Ray, 99051H

\bibitem[{{Gonz{\'a}lez-Caniulef} {et~al.}(2019){Gonz{\'a}lez-Caniulef},
  {Guillot}, \& {Reisenegger}}]{gonzalezcaniulef19}
{Gonz{\'a}lez-Caniulef}, D., {Guillot}, S., \& {Reisenegger}, A. 2019, \mnras,
  490, 5848

\bibitem[{Gough(2009)}]{gough2009gnu}
Gough, B. 2009, GNU scientific library reference manual (Network Theory Ltd.)

\bibitem[{{Gralla} {et~al.}(2017){Gralla}, {Lupsasca}, \&
  {Philippov}}]{Gralla2017}
{Gralla}, S.~E., {Lupsasca}, A., \& {Philippov}, A. 2017, \apj, 851, 137

\bibitem[{{Guillot} {et~al.}(2016){Guillot}, {Kaspi}, {Archibald}, {Bachetti},
  {Flynn}, {Jankowski}, {Bailes}, {Boggs}, {Christensen}, {Craig}, {Hailey},
  {Harrison}, {Stern}, \& {Zhang}}]{guillot16}
{Guillot}, S., {Kaspi}, V.~M., {Archibald}, R.~F., {et~al.} 2016, \mnras, 463,
  2612

\bibitem[{{Haakonsen} {et~al.}(2012){Haakonsen}, {Turner}, {Tacik}, \&
  {Rutledge}}]{haakonsenetal12}
{Haakonsen}, C.~B., {Turner}, M.~L., {Tacik}, N.~A., \& {Rutledge}, R.~E. 2012,
  \apj, 749, 52

\bibitem[{{Hameury} {et~al.}(1983){Hameury}, {Heyvaerts}, \&
  {Bonazzola}}]{hameuryetal83}
{Hameury}, J.~M., {Heyvaerts}, J., \& {Bonazzola}, S. 1983, \aap, 121, 259

\bibitem[{{Handley} {et~al.}(2015){Handley}, {Hobson}, \&
  {Lasenby}}]{2015MNRAS.450L..61H}
{Handley}, W.~J., {Hobson}, M.~P., \& {Lasenby}, A.~N. 2015, \mnras, 450, L61

\bibitem[{{Harding} \& {Muslimov}(2011)}]{2011ApJ...743..181H}
{Harding}, A.~K., \& {Muslimov}, A.~G. 2011, \apj, 743, 181

\bibitem[{{Hebeler} {et~al.}(2013){Hebeler}, {Lattimer}, {Pethick}, \&
  {Schwenk}}]{hebeler13}
{Hebeler}, K., {Lattimer}, J.~M., {Pethick}, C.~J., \& {Schwenk}, A. 2013,
  \apj, 773, 11

\bibitem[{{Heinke}(2013)}]{heinke13}
{Heinke}, C.~O. 2013, Journal of Physics Conference Series, 432, 012001

\bibitem[{{Heinke} {et~al.}(2006){Heinke}, {Rybicki}, {Narayan}, \&
  {Grindlay}}]{heinkeetal06}
{Heinke}, C.~O., {Rybicki}, G.~B., {Narayan}, R., \& {Grindlay}, J.~E. 2006,
  \apj, 644, 1090

\bibitem[{{Heinke} {et~al.}(2014){Heinke}, {Cohn}, {Lugger}, {Webb}, {Ho},
  {Anderson}, {Campana}, {Bogdanov}, {Haggard}, {Cool}, \&
  {Grindlay}}]{2014MNRAS.444..443H}
{Heinke}, C.~O., {Cohn}, H.~N., {Lugger}, P.~M., {et~al.} 2014, \mnras, 444,
  443

\bibitem[{{Ho} \& {Heinke}(2009)}]{hoheinke09}
{Ho}, W.~C.~G., \& {Heinke}, C.~O. 2009, \nat, 462, 71

\bibitem[{{Ho} {et~al.}(2007){Ho}, {Kaplan}, {Chang}, {van Adelsberg}, \&
  {Potekhin}}]{Ho07}
{Ho}, W. C.~G., {Kaplan}, D.~L., {Chang}, P., {van Adelsberg}, M., \&
  {Potekhin}, A.~Y. 2007, \mnras, 375, 821

\bibitem[{{Ho} \& {Lai}(2001)}]{holai01}
{Ho}, W.~C.~G., \& {Lai}, D. 2001, \mnras, 327, 1081

\bibitem[{{Hunter}(2007)}]{4160265}
{Hunter}, J.~D. 2007, Computing in Science Engineering, 9, 90

\bibitem[{{Iglesias} \& {Rogers}(1996)}]{iglesiasrogers96}
{Iglesias}, C.~A., \& {Rogers}, F.~J. 1996, \apj, 464, 943

\bibitem[{{Kalapotharakos} {et~al.}(2019){Kalapotharakos}, {Harding},
  {Kazanas}, \& {Wadiasingh}}]{2019ApJ...883L...4K}
{Kalapotharakos}, C., {Harding}, A.~K., {Kazanas}, D., \& {Wadiasingh}, Z.
  2019, \apjl, 883, L4

\bibitem[{{Kalapotharakos} {et~al.}(2021){Kalapotharakos}, {Wadiasingh},
  {Harding}, \& {Kazanas}}]{2021ApJ...907...63K}
{Kalapotharakos}, C., {Wadiasingh}, Z., {Harding}, A.~K., \& {Kazanas}, D.
  2021, \apj, 907, 63

\bibitem[{{Lai}(2001)}]{lai01}
{Lai}, D. 2001, Reviews of Modern Physics, 73, 629

\bibitem[{{Lattimer} \& {Prakash}(2001)}]{Lattimer01}
{Lattimer}, J.~M., \& {Prakash}, M. 2001, \apj, 550, 426

\bibitem[{{Lattimer} \& {Prakash}(2005)}]{Lattimer05}
---. 2005, Physical Review Letters, 94, 111101

\bibitem[{{Lattimer} \& {Prakash}(2016)}]{lattimer16}
---. 2016, \physrep, 621, 127

\bibitem[{{Lewis}(2019)}]{Lewis2019}
{Lewis}, A. 2019, arXiv e-prints, arXiv:1910.13970

\bibitem[{{Lo} {et~al.}(2013){Lo}, {Miller}, {Bhattacharyya}, \& {Lamb}}]{lo13}
{Lo}, K.~H., {Miller}, M.~C., {Bhattacharyya}, S., \& {Lamb}, F.~K. 2013, \apj,
  776, 19

\bibitem[{{Lockhart} {et~al.}(2019){Lockhart}, {Gralla}, {{\"O}zel}, \&
  {Psaltis}}]{Lockhart2019}
{Lockhart}, W., {Gralla}, S.~E., {{\"O}zel}, F., \& {Psaltis}, D. 2019, \mnras,
  490, 1774

\bibitem[{{Ludlam} {et~al.}(2018){Ludlam}, {Miller}, {Arzoumanian}, {Bult},
  {Cackett}, {Chakrabarty}, {Dauser}, {Enoto}, {Fabian}, {Garc{\'\i}a},
  {Gendreau}, {Guillot}, {Homan}, {Jaisawal}, {Keek}, {La Marr}, {Malacaria},
  {Markwardt}, {Steiner}, \& {Strohmayer}}]{ludlam2018}
{Ludlam}, R.~M., {Miller}, J.~M., {Arzoumanian}, Z., {et~al.} 2018, \apjl, 858,
  L5

\bibitem[{{McClintock} {et~al.}(2004){McClintock}, {Narayan}, \&
  {Rybicki}}]{mcclintocketal04}
{McClintock}, J.~E., {Narayan}, R., \& {Rybicki}, G.~B. 2004, \apj, 615, 402

\bibitem[{{M{\'e}sz{\'a}ros}(1992)}]{meszaros92}
{M{\'e}sz{\'a}ros}, P. 1992, {High-energy radiation from magnetized neutron
  stars.}

\bibitem[{{Mihalas}(1978)}]{mihalas78}
{Mihalas}, D. 1978, {Stellar atmospheres /2nd edition/}

\bibitem[{{Miller}(2013)}]{miller13}
{Miller}, M.~C. 2013, ArXiv e-prints, arXiv:1312.0029

\bibitem[{{Miller} \& {Lamb}(1998)}]{miller98}
{Miller}, M.~C., \& {Lamb}, F.~K. 1998, \apjl, 499, L37

\bibitem[{{Miller} \& {Lamb}(2015)}]{miller15}
---. 2015, \apj, 808, 31

\bibitem[{{Miller} \& {Lamb}(2016)}]{miller16}
---. 2016, European Physical Journal A, 52, 63

\bibitem[{{Miller} {et~al.}(2019){Miller}, {Lamb}, {Dittmann}, B., Z., \&
  C.}]{miller19}
{Miller}, M.~C., {Lamb}, F.~K., {Dittmann}, A.~J., {et~al.} 2019, \apjl, 887,
  L24

\bibitem[{{Morsink} {et~al.}(2007){Morsink}, {Leahy}, {Cadeau}, \&
  {Braga}}]{morsink07}
{Morsink}, S.~M., {Leahy}, D.~A., {Cadeau}, C., \& {Braga}, J. 2007, \apj, 663,
  1244

\bibitem[{{N{\"a}ttil{\"a}} \& {Pihajoki}(2018)}]{2018A&A...615A..50N}
{N{\"a}ttil{\"a}}, J., \& {Pihajoki}, P. 2018, \aap, 615, A50

\bibitem[{{Oertel} {et~al.}(2017){Oertel}, {Hempel}, {Kl{\"a}hn}, \&
  {Typel}}]{Oertel17}
{Oertel}, M., {Hempel}, M., {Kl{\"a}hn}, T., \& {Typel}, S. 2017, Reviews of
  Modern Physics, 89, 015007

\bibitem[{{Oliphant}(2007)}]{4160250}
{Oliphant}, T.~E. 2007, Computing in Science Engineering, 9, 10

\bibitem[{{{\"O}zel}(2013)}]{ozel13}
{{\"O}zel}, F. 2013, Reports on Progress in Physics, 76, 016901

\bibitem[{{{\"O}zel} \& {Psaltis}(2009)}]{Ozel09}
{{\"O}zel}, F., \& {Psaltis}, D. 2009, \prd, 80, 103003

\bibitem[{{Page}(1995)}]{1995ApJ...442..273P}
{Page}, D. 1995, \apj, 442, 273

\bibitem[{{Pavlov} \& {Zavlin}(1997)}]{pavlov97}
{Pavlov}, G.~G., \& {Zavlin}, V.~E. 1997, \apjl, 490, L91

\bibitem[{{Pechenick} {et~al.}(1983){Pechenick}, {Ftaclas}, \&
  {Cohen}}]{pechenick83}
{Pechenick}, K.~R., {Ftaclas}, C., \& {Cohen}, J.~M. 1983, \apj, 274, 846

\bibitem[{{Potekhin}(2010)}]{2010A&A...518A..24P}
{Potekhin}, A.~Y. 2010, \aap, 518, A24

\bibitem[{{Potekhin}(2014)}]{potekhin14}
---. 2014, Physics Uspekhi, 57, 735

\bibitem[{{Poutanen} \& {Beloborodov}(2006)}]{poutanen06}
{Poutanen}, J., \& {Beloborodov}, A.~M. 2006, \mnras, 373, 836

\bibitem[{{Poutanen} \& {Gierli{\'n}ski}(2003)}]{poutanen03}
{Poutanen}, J., \& {Gierli{\'n}ski}, M. 2003, \mnras, 343, 1301

\bibitem[{{Psaltis} \& {{\"O}zel}(2014)}]{psaltis14b}
{Psaltis}, D., \& {{\"O}zel}, F. 2014, \apj, 792, 87

\bibitem[{{Psaltis} {et~al.}(2014){Psaltis}, {{\"O}zel}, \&
  {Chakrabarty}}]{psaltis14a}
{Psaltis}, D., {{\"O}zel}, F., \& {Chakrabarty}, D. 2014, \apj, 787, 136

\bibitem[{{Raaijmakers} {et~al.}(2019){Raaijmakers}, {Riley}, {Watts}, {Greif},
  M., \& K.}]{raaijmakers19}
{Raaijmakers}, G., {Riley}, T.~E., {Watts}, A.~L., {et~al.} 2019, \apjl, 887,
  L22

\bibitem[{{Raaijmakers} {et~al.}(2020){Raaijmakers}, {Greif}, {Riley},
  {Hinderer}, {Hebeler}, {Schwenk}, {Watts}, {Nissanke}, {Guillot}, {Lattimer},
  \& {Ludlam}}]{raaijmakers20}
{Raaijmakers}, G., {Greif}, S.~K., {Riley}, T.~E., {et~al.} 2020, \apjl, 893,
  L21

\bibitem[{{Rajagopal} \& {Romani}(1996)}]{rajagopalromani96}
{Rajagopal}, M., \& {Romani}, R.~W. 1996, \apj, 461, 327

\bibitem[{{Randhawa} {et~al.}(2019){Randhawa}, {Meisel}, {Giuliani}, {Schatz},
  {Meyer}, {Ebinger}, {Hood}, \& {Kanungo}}]{2019ApJ...887..100R}
{Randhawa}, J.~S., {Meisel}, Z., {Giuliani}, S.~A., {et~al.} 2019, \apj, 887,
  100

\bibitem[{{Rauch} {et~al.}(2008){Rauch}, {Suleimanov}, \& {Werner}}]{rauch08}
{Rauch}, T., {Suleimanov}, V., \& {Werner}, K. 2008, \aap, 490, 1127

\bibitem[{{Read} {et~al.}(2009){Read}, {Lackey}, {Owen}, \&
  {Friedman}}]{Read09a}
{Read}, J.~S., {Lackey}, B.~D., {Owen}, B.~J., \& {Friedman}, J.~L. 2009, \prd,
  79, 124032

\bibitem[{{Riley}(2019)}]{Riley19b}
{Riley}, T.~E. 2019, PhD thesis, University of Amsterdam,
  https://hdl.handle.net/11245.1/aa86fcf3-2437-4bc2-810e-cf9f30a98f7a

\bibitem[{{Riley} {et~al.}(2019){Riley}, {Watts}, {Bogdanov}, {Ray}, {Ludlam},
  {Guillot}, {Arzoumanian}, {Baker}, {Bilous}, {Chakrabarty}, {Gendreau},
  {Harding}, {Ho}, {Lattimer}, {Morsink}, \& {Strohmayer}}]{riley19}
{Riley}, T.~E., {Watts}, A.~L., {Bogdanov}, S., {et~al.} 2019, \apjl, 887, L21

\bibitem[{{Romani}(1987)}]{romani87}
{Romani}, R.~W. 1987, \apj, 313, 718

\bibitem[{{Rosen}(1968)}]{Rosen68}
{Rosen}, L.~C. 1968, \apss, 1, 372

\bibitem[{{Salmi} {et~al.}(2020){Salmi}, {Suleimanov}, {N{\"a}ttil{\"a}}, \&
  {Poutanen}}]{Salmi2020}
{Salmi}, T., {Suleimanov}, V.~F., {N{\"a}ttil{\"a}}, J., \& {Poutanen}, J.
  2020, arXiv e-prints, arXiv:2002.11427

\bibitem[{{Salmi} {et~al.}(2019){Salmi}, {Suleimanov}, \& {Poutanen}}]{Salmi19}
{Salmi}, T., {Suleimanov}, V.~F., \& {Poutanen}, J. 2019, \aap, 627, A39

\bibitem[{{Silva} {et~al.}(2020){Silva}, {Pappas}, {Yunes}, \&
  {Yagi}}]{2020arXiv200805565S}
{Silva}, H.~O., {Pappas}, G., {Yunes}, N., \& {Yagi}, K. 2020, arXiv e-prints,
  arXiv:2008.05565

\bibitem[{{Spitkovsky}(2006)}]{2006ApJ...648L..51S}
{Spitkovsky}, A. 2006, \apjl, 648, L51

\bibitem[{{Steiner} {et~al.}(2013){Steiner}, {Lattimer}, \&
  {Brown}}]{2013ApJ...765L...5S}
{Steiner}, A.~W., {Lattimer}, J.~M., \& {Brown}, E.~F. 2013, \apjl, 765, L5

\bibitem[{{Stevens} {et~al.}(2016){Stevens}, {Fiege}, {Leahy}, \&
  {Morsink}}]{2016ApJ...833..244S}
{Stevens}, A.~L., {Fiege}, J.~D., {Leahy}, D.~A., \& {Morsink}, S.~M. 2016,
  \apj, 833, 244

\bibitem[{{Strohmayer}(1992)}]{strohmayer92}
{Strohmayer}, T.~E. 1992, \apj, 388, 138

\bibitem[{{Suleimanov} \& {Werner}(2007)}]{Suleimanov2007}
{Suleimanov}, V., \& {Werner}, K. 2007, \aap, 466, 661

\bibitem[{{Suleimanov} {et~al.}(2014){Suleimanov}, {Klochkov}, {Pavlov}, \&
  {Werner}}]{suleimanovetal14}
{Suleimanov}, V.~F., {Klochkov}, D., {Pavlov}, G.~G., \& {Werner}, K. 2014,
  \apjs, 210, 13

\bibitem[{{Suleimanov} {et~al.}(2010){Suleimanov}, {Pavlov}, \&
  {Werner}}]{2010ApJ...714..630S}
{Suleimanov}, V.~F., {Pavlov}, G.~G., \& {Werner}, K. 2010, \apj, 714, 630

\bibitem[{{Suleimanov} {et~al.}(2012){Suleimanov}, {Pavlov}, \&
  {Werner}}]{2012ApJ...751...15S}
---. 2012, \apj, 751, 15

\bibitem[{{Talts} {et~al.}(2018){Talts}, {Betancourt}, {Simpson}, {Vehtari}, \&
  {Gelman}}]{Talts2018}
{Talts}, S., {Betancourt}, M., {Simpson}, D., {Vehtari}, A., \& {Gelman}, A.
  2018, arXiv e-prints, arXiv:1804.06788

\bibitem[{{Timokhin} \& {Arons}(2013)}]{2013MNRAS.429...20T}
{Timokhin}, A.~N., \& {Arons}, J. 2013, \mnras, 429, 20

\bibitem[{{Tsai}(1974)}]{Tsai74}
{Tsai}, Y.-S. 1974, Reviews of Modern Physics, 46, 815

\bibitem[{{van der Walt} {et~al.}(2011){van der Walt}, {Colbert}, \&
  {Varoquaux}}]{2011CSE....13b..22V}
{van der Walt}, S., {Colbert}, S.~C., \& {Varoquaux}, G. 2011, Computing in
  Science and Engineering, 13, 22

\bibitem[{Virtanen {et~al.}(2020)Virtanen, Gommers, Oliphant, Haberland, Reddy,
  Cournapeau, Burovski, Peterson, Weckesser, Bright, {van der Walt}, Brett,
  Wilson, Millman, Mayorov, Nelson, Jones, Kern, Larson, Carey, Polat, Feng,
  Moore, {VanderPlas}, Laxalde, Perktold, Cimrman, Henriksen, Quintero, Harris,
  Archibald, Ribeiro, Pedregosa, {van Mulbregt}, \& {SciPy 1.0
  Contributors}}]{2020SciPy-NMeth}
Virtanen, P., Gommers, R., Oliphant, T.~E., {et~al.} 2020, Nature Methods, 17,
  261

\bibitem[{{Watts} {et~al.}(2016){Watts}, {Andersson}, {Chakrabarty}, {Feroci},
  {Hebeler}, {Israel}, {Lamb}, {Miller}, {Morsink}, {{\"O}zel}, {Patruno},
  {Poutanen}, {Psaltis}, {Schwenk}, {Steiner}, {Stella}, {Tolos}, \& {van der
  Klis}}]{2016RvMP...88b1001W}
{Watts}, A.~L., {Andersson}, N., {Chakrabarty}, D., {et~al.} 2016, Reviews of
  Modern Physics, 88, 021001

\bibitem[{{Weinberg} {et~al.}(2001){Weinberg}, {Miller}, \&
  {Lamb}}]{weinberg01}
{Weinberg}, N., {Miller}, M.~C., \& {Lamb}, D.~Q. 2001, \apj, 546, 1098

\bibitem[{{Wijngaarden} {et~al.}(2019){Wijngaarden}, {Ho}, {Chang}, {Heinke},
  {Page}, {Beznogov}, \& {Patnaude}}]{wijngaardenetal19}
{Wijngaarden}, M.~J.~P., {Ho}, W.~C.~G., {Chang}, P., {et~al.} 2019, \mnras,
  484, 974

\bibitem[{{Wijngaarden} {et~al.}(2020){Wijngaarden}, {Ho}, {Chang}, {Page},
  {Wijnands}, {Ootes}, {Cumming}, {Degenaar}, \&
  {Beznogov}}]{wijngaardenetal20}
{Wijngaarden}, M.~J.~P., {Ho}, W. C.~G., {Chang}, P., {et~al.} 2020, \mnras,
  493, 4936

\bibitem[{{Wilms} {et~al.}(2000){Wilms}, {Allen}, \&
  {McCray}}]{2000ApJ...542..914W}
{Wilms}, J., {Allen}, A., \& {McCray}, R. 2000, \apj, 542, 914

\bibitem[{{Zavlin} {et~al.}(1996){Zavlin}, {Pavlov}, \&
  {Shibanov}}]{zavlinetal96}
{Zavlin}, V.~E., {Pavlov}, G.~G., \& {Shibanov}, Y.~A. 1996, \aap, 315, 141

\end{thebibliography}

\end{document}